\documentclass[twocolumn]{aa}
\usepackage{graphicx}
\usepackage{longtable}
\usepackage{txfonts}
\begin{document}
   \title{The XMM-{\it Newton} HBS28 sample: studying the
obscuration in hard X-ray selected AGNs
\thanks{Based on observations collected at the European 
Southern Observatory, La Silla, Chile and on observations 
obtained with XMM-{\it Newton}, an ESA science mission
with instruments and contributions directly funded by ESA Member States and 
the USA (NASA)}}
   \subtitle{}
   \author{A. Caccianiga
          \inst{1}
	  \and
	  P. Severgnini\inst{1}
	  \and
         V. Braito\inst{1,2}
         \and
         R. Della Ceca\inst{1}
         \and
         T. Maccacaro\inst{1}
         \and
         A. Wolter\inst{1}
         \and
	 X. Barcons\inst{3}
         \and
         F.J. Carrera\inst{3}
         \and
	 I. Lehmann\inst{4}
	 \and
	 M.J. Page\inst{5}
	 \and
	 R. Saxton\inst{6}
	 \and
	 N. A. Webb\inst{7}
}

   \offprints{caccia@brera.mi.astro.it}

   \institute{INAF-Osservatorio Astronomico di Brera, via Brera 28 - 
               20121 Milano, Italy\\
              \email{caccia, paola, braito, rdc, tommaso, anna@brera.mi.astro.it}     
   \and
     Dipartimento di Astronomia, Universit\`a di  Padova, 
     Vicolo dell'Osservatorio 2, 
     I-35122, Padova, Italy
   \and
     Instituto de Fisica de Cantabria (CSIC-UC), Avenida de los Castros, 
     39005 Santander, Spain\\
     \email{barcons, carreraf@ifca.unican.es}
   \and
      Astrophysikalisches Institut Potsdam (AIP), 
      An der Sternwarte 16, 14482 Potsdam, Germany\\
     \email{ile@xray.mpe.mpg.de}
   \and
      Mullard Space Science Laboratory, University College London, 
      Holmbury St. Mary, Dorking, Surrey, RH5 6NT\\
     \email{mjp@mssl.ucl.ac.uk}
   \and
     X-ray Astronomy Group, Department of Physics and Astronomy, 
     Leicester University, Leicester LE1 7RH, UK\\
    \email{rds@star.le.ac.uk}
   \and
          Centre d'Etude Spatiale des Rayonnements, 9 avenue du Colonel Roche, 
    31028 Toulouse Cedex 04, France\\
     \email{webb@cesr.fr}
             }
   
   \date{}

   \abstract{
This paper presents the analysis of a statistically complete sample of 28 
serendipitous X-ray sources selected in 82 pointed 
XMM-{\it Newton} fields down to a count-rate of 0.002 counts s$^{-1}$
(4.5-7.5 keV energy band). 
This is the first sample selected in this energy range to have complete 
spectroscopic identifications and redshift determinations for all the objects. 
Apart from one Galactic source (an interacting binary), all the objects are AGNs. 
Their optical and X-ray properties (derived from the spectral analysis of the XMM- EPIC 
data) are  compared together. 
The good correlation between the optical spectral type and the
X-ray absorption properties supports the AGN unified model. 
Only one object that does not fit the relation between optical and X-ray absorption 
is found,  namely a Seyfert 1.9  with no evidence of obscuration in the X-ray band 
(N$_H<$1.3$\times$10$^{20}$ cm$^{-2}$). 
In total, 7 sources out of 27 are heavily obscured  in the
X-ray (N$_H>$10$^{22}$ cm$^{-2}$), corresponding to a surface density of 0.7 deg$^{-2}$ at the 
flux limit the sample (4-7$\times$10$^{-14}$ erg s$^{-1}$ cm$^{-2}$ in the 4.5-7.5 keV energy band).
Among these obscured objects, two sources show a large (intrinsic) luminosity (L$_{\rm{[2-10keV]}}>$10$^{44}$ erg s$^{-1}$)
and are thus classified as type~2 QSO.
Finally, we have compared the fraction of X-ray absorbed AGNs 
(26\%) with that predicted by the current XRB synthesis models 
at the flux limit of the survey. We find that the models significantly ($\sim$90\% confidence level) 
over predict the fraction of absorbed AGNs thus confirming also in this hard energy band (4.5-7.5 keV)
similar results recently obtained in the 2-10 keV band.

   }
   \titlerunning{the XMM-{\it Newton} HBS28 sample}
   \maketitle
%

\section{Introduction}

Over 80\% of the cosmic X-ray background (XRB) has been recently resolved 
in the 2-10 keV energy range by deep {\it Chandra} and {\it XMM-Newton} 
observations (Mushotzky et al. 2000; Hasinger et al. 2001; Brandt et al. 2001; 
Rosati et al. 2002; Moretti et al. 2003). Its spectrum can be
successfully reproduced by a combination of obscured and unobscured AGNs
(Setti \& Woltjer 1989; Madau, Ghisellini \& Fabian 1994; 
Comastri et al. 1995; Comastri et al. 2001; Gilli, Salvati \& Hasinger 2001; 
Ueda et al. 2003). The 
obscured AGNs, in particular, play  a
key role in the XRB synthesis models  since they are expected to make
up a significant fraction of the population
in hard X-ray energies (see Fabian 2003 and references therein). 
On the contrary, only a small fraction of absorbed AGNs is 
found in  soft X-ray surveys even at the faintest fluxes 
(e.g. $\sim$15\% in the ROSAT Ultra Deep Survey, 
Lehmann et al. 2001).
 
Many important issues related to this population are still to
be understood, like the number of type~2 QSO, the
relationship between optical absorption and X--ray obscuration
(e.g. Maccacaro, Perola \& Elvis 1982; Panessa \& Bassani 2002; 
Risaliti et al. 1999; Maiolino et al., 2001), the evolutionary properties of 
type~2 AGNs (Gilli, Salvati \& Hasinger 2001; Franceschini et al. 2002) 
and the nature of the ``Optically dull X--ray loud galaxies'', firstly discovered 
from the analysis of {\it Einstein} and ROSAT data (e.g. Elvis et al. 1981; Maccacaro 
et al. 1987; Griffiths et al. 1995; Tananbaum et al. 1997) and more recently
found in large number by Chandra and XMM-{\it Newton} surveys
(Fiore et al., 2000; Barger et al. 2002;  Comastri et al. 2002; 
Severgnini et al. 2003).

Many of these issues can be studied by using 
hard X-ray surveys with complete spectroscopic information,
both in the optical and in the X-rays, to allow a direct
comparison between the two bands (e.g. Akiyama et al. 2003; Ueda et al. 2003). 
In particular, AGN samples selected in the hardest energy band
currently reachable with imaging instruments, i.e. the 5-10 keV or 5-8 keV 
energy band, provide the best starting point for this kind of
analysis since they are less affected by obscuration.

However, the sources found in medium-deep surveys are usually so faint in
the optical that
the completion of the identification process is hard, if not impossible, 
to achieve.
For instance, about 25\% of the sources discovered  in the Chandra Deep Fields 
have optical counterparts fainter than R=25 (Giacconi et al. 2002) 
making the direct redshift estimate very difficult even with the
largest optical telescopes currently available. Furthermore, an
accurate X-ray spectral analysis is usually not feasible for
the faintest sources detected in deep surveys.
Bright X-ray surveys, for which a complete spectroscopic identification in the
optical is 
a feasible task and for which the X-ray spectra  can be easily collected, 
thus complement medium and deep surveys. 

With this goal in mind, the XMM-{\it Newton} Survey Science Centre 
(SSC)\footnote{
The XMM-{\it Newton} SSC is an international collaboration appointed by ESA to 
help the SOC in developing the SAS, to pipe-line process all XMM data
and to exploit 
the XMM-{\it Newton} serendipitous detections. See http://xmmssc-www.star.le.ac.uk/ for
a description of the SSC activities.
}
is building up a large ($\sim 1000$ sources) sample of bright  
(flux limit $\sim$10$^{-13}$ erg cm$^{-2}$ s$^{-1}$)  
serendipitous XMM-{\it Newton} sources at high Galactic latitude ($|b| > 20^o$),
following well defined criteria  so as to allow both a detailed study of sources 
of high individual interest as well as  statistical population  studies
(see Della Ceca et al. 2002; Della Ceca 2002). 

The scope of this paper is to present a first complete and
representative sub-sample of 28 objects, selected in the 4.5-7.5 keV energy 
range. Even if small, this sample has the important advantage 
that a classification based on dedicated optical 
spectroscopy is available for all the sources. In contrast, 
recent samples selected in a similar very hard energy band ($\sim$5-10 keV) 
using BeppoSAX (Fiore et al. 1999; La Franca et al. 2002)
ASCA (Nandra et al. 2003), XMM-{\it Newton} (Baldi et al. 2002; 
Fiore et al. 2003; Mainieri et al. 2002) or Chandra data (e.g. Rosati et al. 2002)
are usually characterized by a partial identification level. 

In Section~2 the sample is presented while in Section~3 and
in Section~4 the optical and the X-ray spectra respectively are discussed. 
In Section~5 the X-ray and optical properties are compared together while in
Section~6 the relative fraction of absorbed versus 
un-absorbed AGNs is compared to the predictions based on the X-ray background 
synthesis models. Finally, in Section~7 an estimate of the density of type~2 QSO 
at the flux limit of the survey is given while in Section~8 we briefly 
discuss the problem of the X-ray bright Optically Normal Galaxies. 
Summary and conclusions are presented in Section~9.
Throughout this paper 
H$_0$=65 km s$^{-1}$ Mpc$^{-1}$, $\Omega_{\Lambda}$=0.7 and $\Omega_{M}$=0.3
are assumed.

\section{The XMM-{\it Newton} Bright Serendipitous Source Sample} 

The XMM-{\it Newton} Bright Serendipitous Source (XMM-{\it Newton} BSS) sample is an
ongoing project aimed at the selection of a statistically 
complete sample of X-ray sources serendipitously discovered in 
pointed XMM-{\it Newton} observations (see Della Ceca et al. 2002, Della
Ceca 2002 for details). The XMM-{\it Newton} BSS consists of 2 complementary samples
based on the XMM-{\it Newton}  EPIC-MOS2 data and 
selected in the 0.5-4.5~keV and in the 4.5-7.5~keV energy range respectively.
The reasons for using the MOS2 detector for the definition of the sample
have been described in details by Della Ceca et al. (2002).
The XMM-{\it Newton} BSS 
will be released to the public community (Della Ceca et al. in prep.) 
and, when combined with medium and
deep X-ray surveys, will allow us to 
investigate on a wide range of luminosities and redshifts 
the properties of the sources responsible for the XRB at energies below 
10 keV.

\subsection{A pilot study: the HBS28 Sample}

As a pilot study we have  selected a representative  
sub-sample (the HBS28 sample) in the 4.5--7.5 keV energy range.
In particular we have used the XMM-{\it Newton} MOS2 fields available to
the XMM-{\it Newton} SSC until
December 2002 and selected according to the criteria
described in Della Ceca et al. (2002) and Della Ceca (2002) i.e.:

\begin{enumerate}

\item $|b^{II}|>$20$^o$

\item Galactic column density (N$_H$) below 10$^{21}$ cm$^{-2}$;

\item Exclusion of fields centered on bright and/or extended
X-ray targets;

\item Exclusion of fields which suffer from very high background flares;

\end{enumerate}

Moreover, to define the pilot sample reported here we have used only
the fields with  $\alpha\leq$ 5h or $\alpha\geq$18h. This last constraint,
coupled with the $|b^{II}|>$20$^o$ criterion, selects an area of sky mostly in 
the Southern hemisphere. 

For the source detection and characterization 
we have followed the same pipeline processing used for the
First XMM-Newton Serendipitous Source Catalogue, which is described in details 
in http://xmmssc-www.star.le.ac.uk/. 
The sources are then selected with the following constraints:

\begin{enumerate}

\item Count rate (4.5-7.5 keV) $\geq$ 0.002 counts s$^{-1}$;

\item maximum likelihood $>$ 12 in the 4.5-7.5 keV band 
(corresponding to a probability for a random 
Poissonian fluctuation to have caused 
the observed source counts of 6$\times$10$^{-6}$);

\item Exclusion of the sources too close to the gaps between the CCDs 
(see Della Ceca et al., in prep. for more details). 

\item Exclusion of the target of the observation 
and of the sources too close to the edge of the field of view. To this
end, we have defined an inner ($r_i$) and an outer ($r_o$) radius
which define the portion of each field used to select the 
sources.

\end{enumerate}

In total, the HBS28 sample contains 28 sources selected in 
82 XMM-{\it Newton}/ EPIC-MOS2 fields. The complete list
of the 82 XMM-{\it Newton} fields  is reported in Table~\ref{fields}.
The count-rate limit of 0.002 counts s$^{-1}$ corresponds to a
nominal flux limit of about 7$\times$10$^{-14}$ erg s$^{-1}$ cm$^{-2}$ 
in the 4.5-7.5 keV energy band for a power-law spectrum with
a photon index $\Gamma$=2. We note, however, that in some object 
the 4.5-7.5 keV flux computed on the basis of the X-ray spectral
analysis (see Section~4) are lower (up to a factor $\sim$2) 
than this formal limit. Therefore, the actual flux limit of the
sample is between 4 and 7$\times$10$^{-14}$ erg s$^{-1}$ cm$^{-2}$.

The total area covered by the HBS28 sample is 9.756 Deg$^2$.
The list of 28 sources selected in the sample is presented in Table~\ref{xraydata}.

\begin{table*}
\caption{The XMM-{\it Newton} fields used for the selection of the HBS28 sample}

\label{fields}
\begin{tabular}{c c r r l ||c c r r l }
          \noalign{\smallskip}
          \hline
obsid & position (J2000) & t & r$_i$ & r$_o$ & obsid & position (J2000) & t & r$_I$ & r$_o$ \\ 
      &                  & (sec)         & ($^\prime$)  &  ($^\prime$) &       &                  & (sec)         & ($^\prime$)  &  ($^\prime$)\\
            \noalign{\smallskip}
            \hline

0125310101 & 00 00 30.4 $-$25 06 43.4 &    19163.6 &          1 &         13   &  0067340101 & 18 55 37.1 $-$46 30 57.6 &    10653.3 &  	0 &	    11  \\
0101040101 & 00 06 19.7 +20 12 22.8 &    34046.5 &          8 &         13  	  &0081341001 & 19 31 21.7 $-$72 39 13.3 &    15288.1 &  	1 &	    13  \\
0127110201 & 00 10 31.2 +10 58 40.7 &     7558.3 &          2 &         13  	  &0081340501 & 20 13 30.0 $-$41 47 25.7 &    19196.8 &  	0 &	    13  \\
0111000101 & 00 18 33.0 +16 26 08.0 &    31363.7 &          3 &         13  	  &0111180201 & 20 40 10.0 $-$00 52 14.7 &    16495.8 &  	8 &	    13  \\
0001930101 & 00 26 06.8 +10 41 12.5 &    18033.7 &          1 &         13  	  &0111510101 & 20 41 50.9 $-$32 26 19.8 &    15046.1 &  	8 &	    13  \\
0050140201 & 00 26 35.9 +17 09 37.8 &    50383.0 &          3 &         13  	  &0130720201 & 20 44 09.0 $-$10 43 11.2 &     5312.0 &  	8 &	    13  \\
0065770101 & 00 32 47.0 +39 34 33.3 &     7360.3 &          1 &         13  	  &0111420101 & 20 45 09.4 $-$31 20 36.6 &    43323.5 &  	8 &	    13  \\
0125320701 & 00 50 03.0 $-$52 07 36.7 &    16525.7 &          1 &         13   &  0112600501 & 20 46 20.0 $-$02 48 48.4 &    10550.3 &  	8 &	    13  \\
0110890401 & 00 57 20.0 $-$22 23 04.1 &    29743.3 &          8 &         13   &  0083210101 & 20 54 18.8 $-$15 55 38.2 &    10435.5 &  	2 &	    13  \\
0112650401 & 01 04 24.0 $-$06 24 10.6 &    23697.1 &          0 &         13   &  0112190601 & 20 56 21.6 $-$04 37 59.4 &    16639.4 &  	2 &	    13  \\
0103861601 & 01 05 16.4 $-$58 26 10.4 &     7127.5 &          8 &         13   &  0081340401 & 20 58 27.1 $-$42 38 56.9 &    15002.3 &  	1 &	    13  \\
0101040201 & 01 23 46.0 $-$58 48 25.8 &    28995.6 &          8 &         13   &  0041150101 & 21 04 11.1 $-$11 21 40.0 &    38736.2 &  	1 &	    13  \\
0109860101 & 01 25 33.4 +01 45 38.2 &    38535.6 &          4 &         13  	  &0038540301 & 21 04 40.2 $-$12 20 05.6 &    14698.6 &  	2 &	    13  \\
0112600601 & 01 27 32.1 +19 10 32.3 &     3994.4 &          8 &         13  	  &0088020201 & 21 27 38.1 $-$44 48 38.5 &    16094.7 &  	1 &	    13  \\
0084230301 & 01 31 53.7 $-$13 36 56.4 &    20447.4 &          8 &         13   &  0103060101 & 21 29 12.2 $-$15 38 34.7 &    22042.9 &  	3 &	    13  \\
0032140401 & 01 40 12.1 $-$67 48 40.3 &     7611.1 &          0 &         13   &  0109463501 & 21 37 56.5 $-$43 42 19.8 &     7577.7 &  	2 &	    13  \\
0093641001 & 01 43 01.7 +13 38 24.4 &     9329.0 &          4 &         13  	  &0061940201 & 21 38 07.9 $-$42 36 06.6 &     4868.2 &  	1 &	    13  \\
0101640201 & 01 59 50.2 +00 23 46.8 &     7543.7 &          8 &         13  	  &0008830101 & 21 40 15.5 $-$23 39 32.1 &    13986.7 &  	2 &	    13  \\
0084140101 & 02 08 38.2 +35 23 00.2 &    38003.5 &          2 &         13  	  &0103060401 & 21 51 55.9 $-$30 27 43.5 &    24072.9 &  	3 &	    13  \\
0112371701 & 02 17 12.4 $-$04 39 04.2 &    20002.1 &          0 &         11   &  0124930201 & 21 58 52.9 $-$30 13 28.8 &    36246.3 &  	8 &	    13  \\
0112371001 & 02 18 00.3 $-$04 59 47.8 &    50802.9 &          0 &         11   &  0130920101 & 22 03 09.3 +18 52 27.9 &    16491.3 &	      1 &	  13  \\
0112371501 & 02 18 48.3 $-$04 39 02.9 &     6684.6 &          0 &         11   &  0012440301 & 22 05 10.1 $-$01 55 11.2 &    30937.5 &  	2 &	    13  \\
0112370301 & 02 19 36.3 $-$04 59 47.7 &    50386.7 &          0 &         11   &  0106660101 & 22 15 31.9 $-$17 44 02.6 &    56911.8 &  	1 &	    13  \\
0098810101 & 02 36 12.3 $-$52 19 55.6 &    23451.5 &          2 &         13   &  0009650201 & 22 17 55.4 $-$08 20 58.0 &    21027.5 &  	8 &	    13  \\
0075940301 & 02 36 57.9 +24 38 53.7 &    47428.9 &          8 &         13  	  &0049340201 & 22 20 45.1 $-$24 40 58.1 &    26842.3 &  	6 &	    13  \\
0067190101 & 02 38 19.4 $-$52 11 34.1 &    25628.7 &          3 &         10   &  0018741701 & 22 34 32.9 $-$37 43 48.1 &     7080.3 &  	4 &	    13  \\
0111200101 & 02 42 41.1 $-$00 00 53.7 &    35486.8 &          8 &         13   &  0111790101 & 22 35 45.9 $-$26 03 00.4 &    41304.4 &  	4 &	    13  \\
0111490401 & 02 48 43.8 +31 06 59.2 &    31030.1 &          8 &         13  	  &0103860201 & 22 36 55.9 $-$22 13 10.1 &     8489.9 &  	8 &	    13  \\
0056020301 & 02 56 32.8 +00 06 01.7 &    17251.4 &          2 &         13  	  &0006810101 & 22 42 39.5 +29 43 35.6 &     7009.4 &	      8 &	  13  \\
0041170101 & 03 02 38.5 +00 07 31.9 &    47278.1 &          0 &         13  	  &0109070401 & 22 48 41.5 $-$51 09 57.9 &    14722.6 &  	8 &	    13  \\
0042340501 & 03 07 03.8 $-$28 40 24.1 &    13285.7 &          4 &         13   &  0112240101 & 22 49 48.3 $-$64 23 11.2 &    30608.4 &  	8 &	    13  \\
0122520201 & 03 11 59.3 $-$76 51 53.0 &    28296.4 &          2 &         13   &  0081340901 & 22 51 49.4 $-$17 52 17.0 &    22398.6 &  	1 &	    13  \\
0110970101 & 03 13 09.6 $-$55 03 48.4 &    10240.0 &          0 &         13   &  0112910301 & 22 53 58.8 $-$17 33 55.8 &     5301.3 &  	8 &	    13  \\
0105660101 & 03 17 56.0 $-$44 14 15.1 &    23151.9 &          6 &         13   &  0112170301 & 23 03 15.8 +08 52 25.9 &    23349.6 &	      4 &	  13  \\
0108060501 & 03 32 29.1 $-$27 48 27.1 &    46835.8 &          0 &         13   &  0109130701 & 23 04 43.6 $-$08 41 14.5 &    10615.6 &  	8 &	    13  \\
0099010101 & 03 35 27.6 $-$25 44 54.5 &    18777.9 &          8 &         13   &  0033541001 & 23 04 45.0 +03 11 35.6 &    12461.5 &	      2 &	  13  \\
0055140101 & 03 39 35.0 $-$35 25 58.1 &    47637.0 &          1 &         11   &  0123900101 & 23 13 58.9 $-$42 43 28.3 &    37401.3 &  	6 &	    13  \\
0111970301 & 04 09 07.2 $-$71 17 43.9 &    18741.3 &          8 &         13   &  0109463601 & 23 15 18.7 $-$59 10 31.7 &     5403.6 &  	8 &	    13  \\
0112600401 & 04 25 44.2 $-$57 13 34.4 &     7627.3 &          8 &         13   &  0112880301 & 23 31 49.9 +19 56 28.5 &    14318.0 &	      3 &	  13  \\
0103861701 & 04 35 17.1 $-$78 01 54.3 &     8010.9 &          8 &         13   &  0093550401 & 23 33 40.0 $-$15 17 12.2 &    22258.0 &  	1 &	    13  \\
0112880401 & 04 59 35.4 +01 47 16.0 &    18915.3 &          2 &         13  	  &0100241001 & 23 49 40.6 +36 25 30.6 &     8739.0 &	      3 &	  13  \\

           \noalign{\smallskip}
            \hline
\end{tabular}

column 1: observation ID; column 2: sky position of the center of the field; column 3: MOS2 exposure time 
(after having removed  the time intervals characterized 
by a high background);
column 4: inner radius (see text for details); 
column 5: outer radius (see text for details).

\end{table*}
\normalsize


\begin{table*}
\caption{The HBS28 sample}
\label{xraydata}
\begin{tabular}{c l c r c l}
          \noalign{\smallskip}
          \hline
name & Observation ID & Gal. N$_H$ & CR$_{4.5-7.5 keV}$ &  EPIC cameras &  Notes\\ 
     &                & ($\times$10$^{20}$) &  ($\times$10$^{-3}$) &             &       \\
     &                & [cm$^{-2}$]   &  [s$^{-1}$]          &              &       \\
            \noalign{\smallskip}
            \hline
\object{XBSJ002618.5$+$105019}                            &   0001930101 &         5.07 &     2.3 &   MOS2,MOS1,pn  &	     \\   
\object{XBSJ013240.1$-$133307}                            &   0084230301 &         1.64 &     3.2 &	MOS2,MOS1,pn  &	      \\   
\object{XBSJ013944.0$-$674909}                            &   0032140401 &         2.49 &     2.0 &	MOS2,MOS1,pn  &	      \\   
\object{XBSJ014100.6$-$675328}                            &   0032140401 &         2.49 &    70.9 &	MOS2,MOS1	   &   a      \\   
\object{XBSJ015957.5$+$003309}                            &   0101640201 &         2.59 &     3.8 &	MOS2,MOS1,pn  &	      \\   
\object{XBSJ021640.7$-$044404}                            &   0112371701 &         2.42 &     2.1 &	 MOS2,MOS1,pn  &	       \\   
\object{XBSJ021808.3$-$045845}                            &   0112371001$^*$ &     2.52 &     2.7 &    MOS2,MOS1,pn  &	  \\   
\object{XBSJ021817.4$-$045113}                            &   0112371001$^*$ &     2.52 &     3.3 &    MOS2,MOS1,pn  &	 \\   
\object{XBSJ021822.2$-$050615}                            &   0112371001$^*$ &     2.52 &     4.5 &     MOS2,MOS1,pn	&	   \\	
\object{XBSJ023713.5$-$522734}                            &   0098810101 &         3.28 &     3.2 &	MOS2,MOS1,pn  &	      \\   
\object{XBSJ030206.8$-$000121}                            &   0041170101 &         7.16 &     3.1 &	MOS2,MOS1,pn  &	      \\   
\object{XBSJ030614.1$-$284019}                            &   0042340501 &         1.36 &     4.6 &	  MOS2,MOS1      &   e	\\   
\object{XBSJ031015.5$-$765131}                            &   0122520201 &         8.21 &     4.4 &	MOS2,MOS1,pn  &	      \\   
\object{XBSJ031146.1$-$550702}                            &   0110970101 &         2.55 &     5.9 &	MOS2,MOS1,pn  &	      \\   
\object{XBSJ031859.2$-$441627}                            &   0105660101 &         2.60 &     2.2 &	MOS2,pn	   &   c      \\   
\object{XBSJ033845.7$-$352253}                            &   0055140101 &         1.31 &     2.4 &	MOS2,MOS1,pn  &	      \\   
\object{XBSJ040658.8$-$712457}                            &   0111970301 &         7.57 &     3.4 &	MOS2,MOS1,pn  &   b      \\   
\object{XBSJ040758.9$-$712833}                            &   0111970301 &         7.57 &     5.0 &	MOS2,pn	   &   c      \\   
\object{XBSJ041108.1$-$711341}                            &   0111970301 &         7.57 &     2.2 &	MOS2,MOS1	   &   a,b    \\   
\object{XBSJ185613.7$-$462239}                            &   0067340101 &         5.29 &     8.2 &	MOS2,MOS1,pn  &	     \\   
\object{XBSJ193248.8$-$723355}                            &   0081341001 &         5.95 &     4.5 &	MOS2,MOS1,pn  &	    \\   
\object{XBSJ204043.4$-$004548}                            &   0111180201 &         6.72 &     3.2 &	MOS2,MOS1,pn  &	      \\   
\object{XBSJ205635.7$-$044717}                            &   0112190601 &         4.96 &     2.1 &	MOS2,MOS1	   &   e      \\   
\object{XBSJ205829.9$-$423634}                            &   0081340401 &         3.89 &     3.9 &	MOS2,MOS1,pn  &	      \\   
\object{XBSJ213002.3$-$153414}                            &   0103060101 &         4.99 &     2.3 &	MOS2,MOS1	   &   d      \\   
\object{XBSJ213824.0$-$423019}                            &   0061940201 &         2.68 &     2.5 &	MOS2,MOS1	   &   e      \\   
\object{XBSJ214041.4$-$234720}                            &   0008830101 &         3.55 &     3.3 &	MOS2,MOS1,pn  &	    \\   
\object{XBSJ220601.5$-$015346}                            &   0012440301 &         6.13 &     2.3 &	MOS2,MOS1	   &   d      \\

            \noalign{\smallskip}
            \hline
\end{tabular}

column 1: source name;

column 2: identification number of the corresponding XMM observation;

column 3: Galactic column density;

column 4: source count-rate in the 4.5-7.5 keV energy band;

column 5: EPIC cameras used for the spectral analysis (see notes);

column 6: notes. 

a = source on a gap in the pn 

b = MOS1 and MOS2 not summed

c = source on a gap in MOS1

d = source outside the pn field of view

e = problems with the pn data

$^{*}$  for the spectral analysis we have summed these data with the 
0112370101 sequence

\end{table*}

\section{Optical identification and classification}

\subsection{The optical counterparts}
The optical counterpart for each source in the sample has been searched
in the Digitized Sky Survey (DSS) and a magnitude has been collected
from the APM (Automatic Plate Measuring machine) 
catalogue\footnote{http://www.ast.cam.ac.uk/$\sim$apmcat/}. Thanks to 
the good positional accuracy\footnote{
see the ``EPIC status of calibration and data analysis'' report, at 
xmm.vilspa.esa.es/} ($\sim$3$\arcsec$-4$\arcsec$) and the brightness of the 
XMM sources, there is no ambiguity in the
detection of  the optical counterpart. When another source is present
just outside the X-ray error circle we have spectroscopically 
observed also this second object. 
In all these cases (4 in total) the object within the error circle 
has turned out to be a more reliable counterpart (an AGN) than the
object outside the circle (always a star) thus confirming the good accuracy
of the X-ray position. All the optical counterparts are within 
4$\arcsec$ from the X-ray position.

\begin{figure}
   \centering
   \includegraphics[width=9cm]{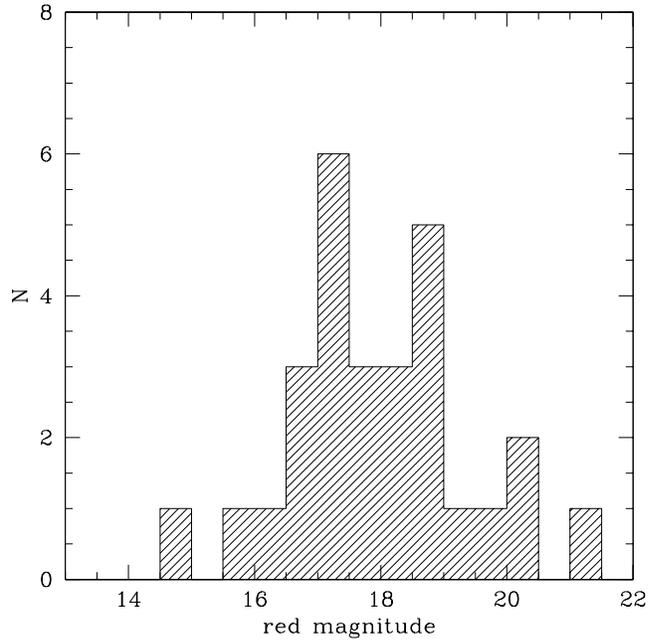}
      \caption{Red magnitude distribution of the 28 sources
in the HBS28 sample. For most of the object (26) the magnitude is
the APM red magnitude, while for the 2 faintest sources the R magnitude 
is considered.
              }
         \label{mag}
   \end{figure}

For 26 sources out of 28, an APM counterpart with a red magnitude brighter than
20.1 was found. For the remaining 2 sources, an optical counterpart is
visible in the acquisition image taken at the ESO 3.6m telescope during
the spectroscopic run (see below for the details) and a magnitude (R) 
is computed.
Hence, for the totality of the sample an optical counterpart has been found. 
The magnitude distribution of the 28 sources is presented in Figure~\ref{mag}.

It is worth noting that, given the small X-ray-to-optical positional offset
and the brightness of the optical counterparts, the expected number of 
spurious optical identifications is completely negligible in the HBS28 sample. 
Even
considering the stars, that represent the most probable chance coincidence,
the expected number of spurious matches in the HBS28 sample 
is below unity. Indeed, as explained below, no ``normal'' stars
are found in the HBS28 sample. The only Galactic object found is
an accreting binary star whose X-ray emission properties are 
well known.
%

   \begin{figure}
   \centering
   \includegraphics[width=9cm]{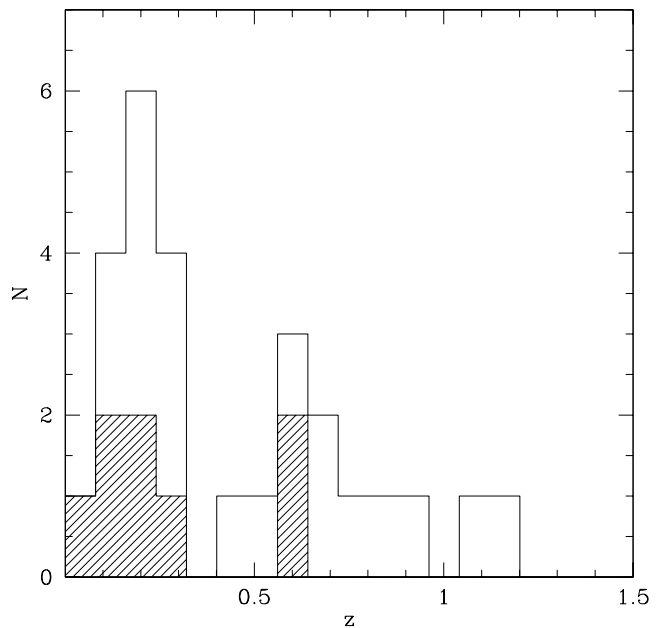}
      \caption{Redshift distribution of the 27 extragalactic
sources in the HBS28 sample. The shaded area represents the type~2 
objects as defined in the text.
              }
         \label{z}
   \end{figure}

\subsection{The optical spectrosocopy}
A search through the literature has produced 5 spectroscopic
identifications (3 QSOs, 1 Narrow Line Radio Galaxy and 1 accreting binary star). Three more
identifications have been taken from the AXIS project  
(Barcons et al. 2002a,b); one of these 3 objects has been re-observed at the SUBARU telescope 
(Severgnini et al. 2003). 
The remaining sources  have been observed and spectroscopically 
identified during dedicated 
observing runs. Most of the sources were observed 
at the ESO 3.6m in May and October 2002 using 
the EFOSC2 coupled with CCD40 and grism n.13 (October) or  grism n.6 
(May). The slit width ranges 
from 1.2\arcsec up to 1.5\arcsec, depending on the seeing conditions,
and was usually oriented according to the parallactic angle to avoid
differential flux losses, except when it was oriented so as to
include two  objects. 
Four additional objects were observed at the TNG telescope in
November 2001 and September 2002, using DOLORES and grism LR-B.
For the data reduction we have used the standard IRAF {\it longslit}
package. The spectra have been wavelength calibrated using an 
Helium-Argon reference spectrum. A relative flux calibration has been
obtained by observing the standard stars G191-B2B, BD25+4655 (TNG) and 
LTT 7987 (ESO). The weather conditions (not always
photometric) and the observing set-up usually do not  guarantee
a correct absolute flux calibration. For this reason, the 
spectra reported here are given in arbitrary units.

\subsection{The classification criteria}

All the 28 sources selected in the HBS28 sample are emission line objects, 
including one accreting binary star (BL Hyi) which is the only Galactic source in 
the sample. We will not discuss  this object any further as the main
scientific target of the HBS28 is the extragalactic population. In Appendix (A.1) 
we briefly outline the main properties of this source.
 
A redshift has been secured for all the 27 extragalactic objects 
in the sample. The redshift distribution is shown in Figure~\ref{z}.

To classify the emission line objects we have adopted the
criteria presented for instance in V\'eron-Cetty \& V\'eron (2001) 
which are based on the line widths
and the [OIII]$\lambda$5007\AA/H$\beta$ flux ratio. 
In particular, we have divided the extragalactic sources into two groups,
i.e. type~1 and type~2 objects:

   \begin{figure*}
   \centering
   \includegraphics[width=18cm]{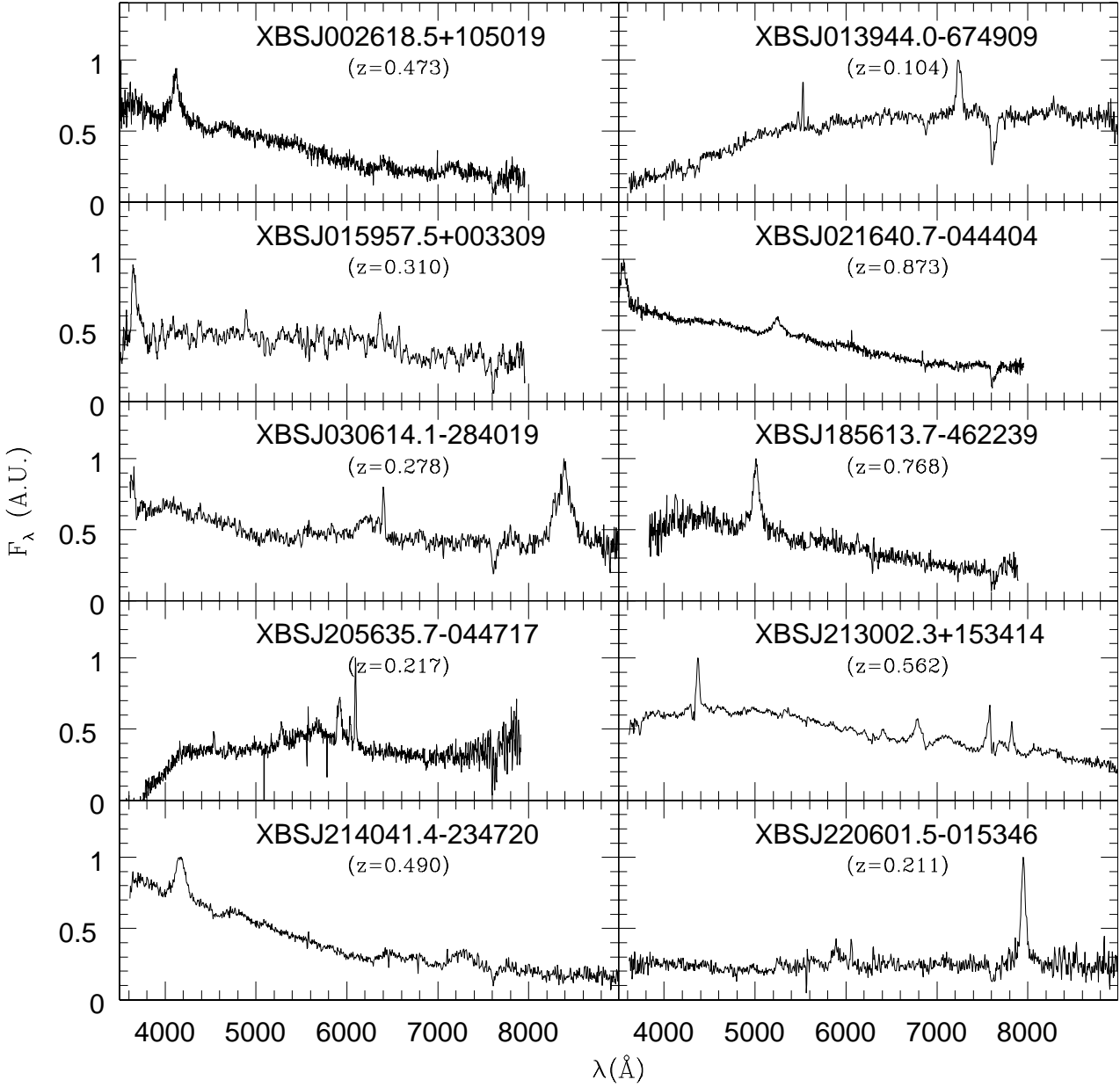}
      \caption{The optical spectra of 10  newly discovered  
Broad Emission Lines AGNs (FWHM$>$2000 km s$^{-1}$) 
              }
         \label{type1}
   \end{figure*}
%

\begin{figure}
   \centering
   \includegraphics[width=9cm]{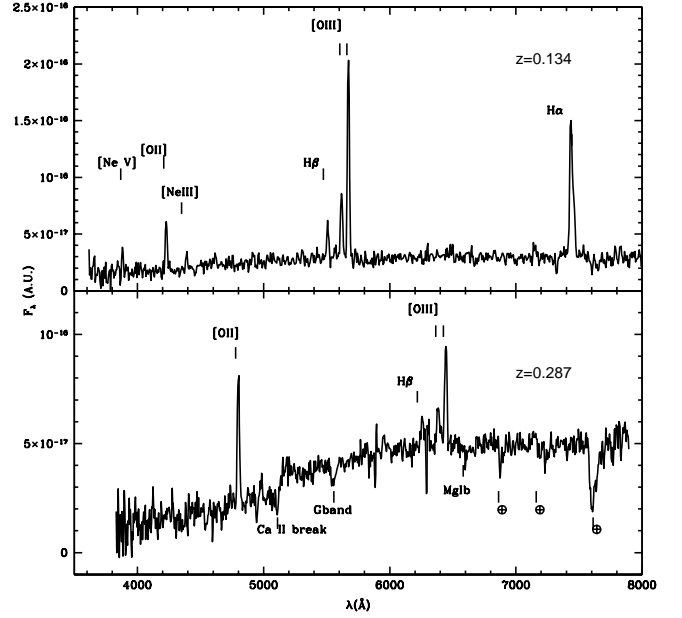}
      \caption{{\it Top panel}: the spectrum of the NEL AGN (a Seyfert 2)  
XBSJ040758.9--712833. {\it Bottom panel}: The spectrum of the object XBSJ193248.8--723355 
optically classified as a Emission Line Galaxy (ELG, i.e.a starburst or a LINER, 
see text for details). As explained in the
text, the X-ray analysis indicates the presence of an AGN in this object. 
              }
         \label{nelg_sy2}
   \end{figure}

\begin{figure}
   \centering
   \includegraphics[width=9cm]{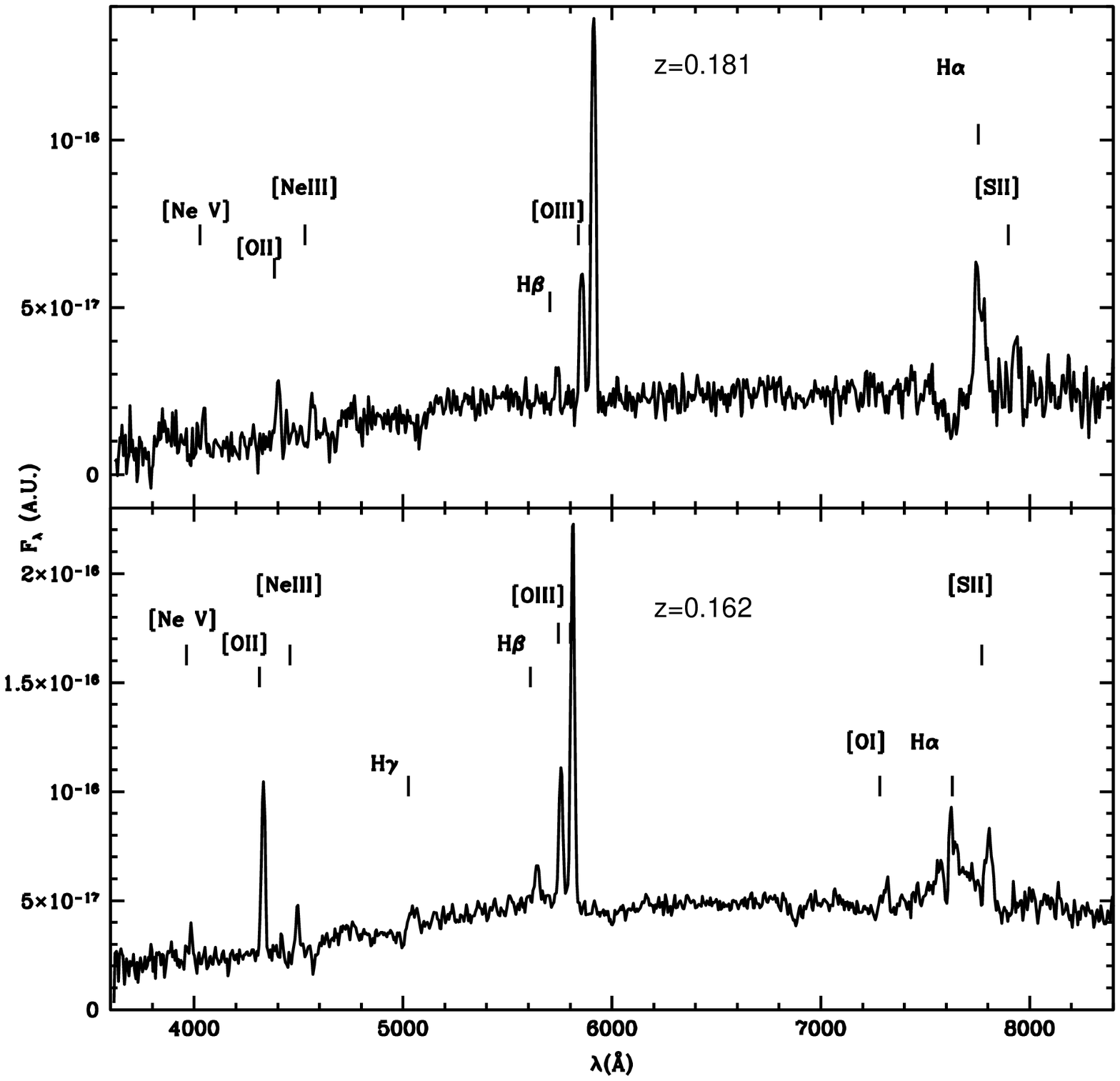}
      \caption{The spectra of the two Sy1.9 discovered in the 
sample: XBSJ040658.8--712457 (top) and XBSJ031146.1--550702 (bottom).
              }
         \label{sy19}
   \end{figure}

\begin{itemize}

\item {\bf Type~1 objects}, including:

\subitem 17 sources showing broad (FWHM$>$2000 kms$^{-1}$) permitted lines. 
These are Broad Emission Line AGNs (BEL AGN). This class includes Seyfert~1,
Broad Emission Line Radio galaxies (BLRG) and type1 QSO. The 
optical spectra of the 10 BEL AGNs identified by our own 
spectroscopic observations are shown in Figure~\ref{type1}. 

\subitem 2 sources which show permitted lines with 
1000 km s$^{-1}$ $<$FWHM$<$2000 km s$^{-1}$ and 
a [OIII]$\lambda$5007/H$\beta$ ratio below 3. These sources  
are classified as  Narrow Line Seyfert 1 (NLSy1) candidates.
The presence of a relatively strong
FeII$\lambda$4570\AA\ hump  further supports the classification of 
these sources as NLSy1.
The optical spectra of the 2 NLSy1 are plotted in Figure~\ref{nlsy1}. 
Given the low-resolution ($\sim$20\AA) and the relatively low S/N ratio
of these spectra, a more detailed analysis of the line ratios
(e.g. the FeII$\lambda$4570\AA/H$\beta$ used for a more quantitative
separation between NLSy1 and Sy1 e.g. by V\'eron-Cetty, V\'eron \& Gon\c{c}alves
2001) is not possible.

   \begin{figure}
   \centering
   \includegraphics[width=9cm]{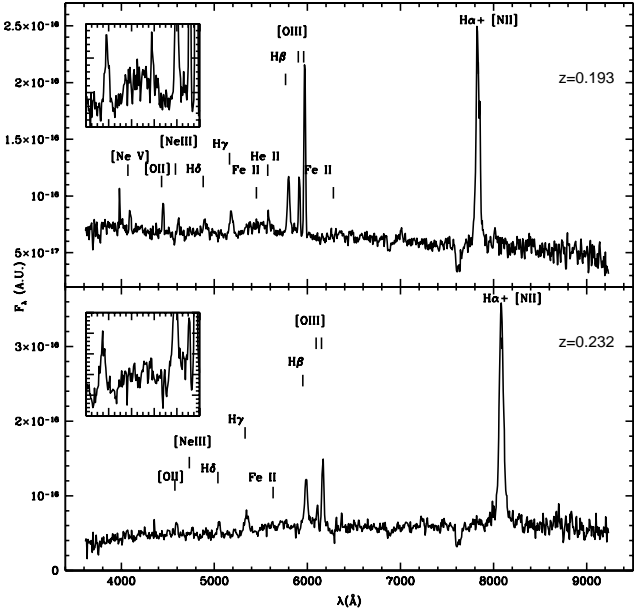}
      \caption{The spectra of the 2 NLSy1 candidates discovered in the 
sample: XBSJ023713.5--522734 (top) and XBSJ205829.9--423634 (bottom).
The two small boxes inside the two figures show the expanded 
region around the Fe~II$\lambda$4600\AA\ hump. 
              }
         \label{nlsy1}
   \end{figure}
%
\item {\bf Type~2 objects}, including:

\subitem 5 sources which show only narrow lines (FWHM$<$1000 km s$^{-1}$) 
and [OIII]$\lambda$5007/H$\beta$$>$3. These objects  are classified as 
Narrow Emission Line AGNs. This class includes Seyfert~2 galaxies, 
Narrow Line Radio galaxies and type~2 QSO. The optical spectrum 
of the Seyfert~2 identified by us is shown in Figure~\ref{nelg_sy2} (top panel);

\subitem 1 object (XBSJ193248.8--723355) whose spectrum presents only narrow lines but the 
[OIII]$\lambda$5007/H$\beta$ ratio is about 2 thus suggesting a non-AGN origin. 
Also the [OII]$\lambda$3727\AA/[OIII]$\lambda$5007 ratio is quite high ($\sim$3)
in agreement with what is usually observed in starburst or HII-region 
galaxies (e.g. Veilleux \& Osterbrock 1987). 
This source is thus classified as an Emission Line 
Galaxy (ELG); its spectrum is shown in Figure~\ref{nelg_sy2} (bottom panel).
As described in the next Section, the X-ray data clearly reveal the presence
of an obscured AGN in this object;

\subitem 2 sources in which all the observed emission lines 
are narrow (FWHM$<$1000 km s$^{-1}$) except for a broad H$\alpha$ 
component. These sources are also characterized by a
high [OIII]$\lambda$5007/H$\beta$ ratio ($>$3) and 
are classified as Seyfert 1.9 (Sy1.9). 
The optical spectra of the two Sy1.9 are reported in Figure~\ref{sy19}.
In one of the sources (XBSJ031146.1--550702) the broad  H$\alpha$ 
component  is clearly evident while in 
XBSJ040658.8--712457 its presence is less obvious and it would require a 
proper de-blending to be clearly quantified.

\end{itemize}

%

In Table~\ref{summary} the optical properties of the 28 sources in the HBS28 sample 
are presented
while the breakdown of the optical classification is summarized in 
Table~\ref{id}. 

In conclusion,  the percentage of type~1 and type~2 objects 
is 70\% and 30\% respectively of the extragalactic sources.
The redshift distributions for type~1 and type~2 objects 
taken separately are only marginally different 
(K-S probability of 6\% of being drawn from the same parent population).

%

\begin{table*}
\caption{The optical properties of HBS28}
\label{summary}
\begin{tabular}{c c c c r l l c}
          \noalign{\smallskip}
          \hline
name & X-ray position & optical position & id & type & z & mag & id ref \\ 
     &  (J2000) & (J2000) & & & & & \\
            \noalign{\smallskip}
            \hline

XBSJ002618.5+105019 & 00 26 18.50 +10 50 19.3 & 00 26 18.70 +10 50 19.5 &  BEL AGN   & 1 &0.473 & 17.5 &   TNG11/01   \\
XBSJ013240.1--133307 & 01 32 40.11 --13 33 07.8 &  01 32 40.27 --13 33 07.2 & NEL AGN   & 2 & 0.562 & 20.0  & ESO10/02   \\
XBSJ013944.0--674909 & 01 39 44.02 --67 49 09.4 &  01 39 43.77 --67 49 08.4 & BEL AGN  & 1 &0.104 & 16.7 &   ESO10/02   \\
XBSJ014100.6--675328 & 01 41 00.66 --67 53 28.9&  01 41 00.23 --67 53 27.6 & E.L.star & -- & -- & 16.4 &  (a) \\
XBSJ015957.5+003309 & 01 59 57.52 +00 33 09.7&  01 59 57.60 +00 33 10.5 & BEL AGN   & 1 &0.310 & 18.3 &   TNG11/01   \\
XBSJ021640.7--044404 & 02 16 40.74 --04 44 04.9&  02 16 40.66 --04 44 04.7 & BEL AGN   & 1 &0.873 & 16.5 &  TNG11/01   \\
XBSJ021808.3--045845 & 02 18 08.34 --04 58 45.7&  02 18 08.20 --04 58 45.7 & BEL AGN   & 1 &0.712 & 17.7 &  AXIS \\
XBSJ021817.4--045113 & 02 18 17.40 --04 51 13.3 &  02 18 17.43 --04 51 13.0& BEL AGN   & 1 &1.070 & 19.5 &   AXIS \\
XBSJ021822.2--050615 & 02 18 22.28 --05 06 15.7&  02 18 22.15 --05 06 14.5 & NEL AGN   & 2 &0.044 & 14.8 & (b) \\
XBSJ023713.5--522734 & 02 37 13.54 --52 27 34.4&  02 37 13.66 --52 27 34.7 & NLSy1  & 1 &0.193 & 17.1 &  ESO10/02   \\
XBSJ030206.8--000121 & 03 02 06.86 --00 01 21.2&  03 02 06.76 --00 01 21.7 & BEL AGN   & 1 &0.641 & 18.8 &  (c) \\
XBSJ030614.1--284019 & 03 06 14.19 --28 40 19.9&  03 06 14.21 --28 40 19.3 & BEL AGN   & 1 &0.278 & 18.5 &  ESO10/02   \\
XBSJ031015.5--765131 & 03 10 15.56 --76 51 31.5&  03 10 15.88 --76 51 33.8 & BEL AGN   & 1 &1.187 & 17.6 &  (d) \\
XBSJ031146.1--550702 & 03 11 46.12 --55 07 02.5&  03 11 46.18 --55 06 59.9 & Sy1.9  & 2 &0.162 & 17.3 & ESO10/02   \\
XBSJ031859.2--441627 & 03 18 59.29 --44 16 27.6&  03 18 59.41 --44 16 26.7 & BEL AGN  & 1 &0.140 & 15.9 &  (b), ESO10/02   \\
XBSJ033845.7--352253 & 03 38 45.77 --35 22 53.4&  03 38 46.02 --35 22 52.7 & NEL AGN   & 2 &0.113 & 17.0 &  (e) \\
XBSJ040658.8--712457 & 04 06 58.87 --71 24 57.7&  04 06 58.88 --71 24 59.8 & Sy1.9  & 2 &0.181 & 18.7 &  ESO10/02   \\
XBSJ040758.9--712833 & 04 07 58.97 --71 28 33.5&  04 07 58.58 --71 28 32.2 & NEL AGN   & 2 &0.134 & 17.0 & ESO10/02   \\
XBSJ041108.1--711341 & 04 11 08.10 --71 13 41.1&  04 11 08.78 --71 13 42.8 & BEL AGN  & 1 &0.923? & 20.3$^*$ & ESO10/02   \\
XBSJ185613.7--462239 & 18 56 13.75 --46 22 39.2 &  18 56 13.78 --46 22 37.1& BEL AGN   & 1 &0.768 & 19.0 & ESO05/02   \\
XBSJ193248.8--723355 & 19 32 48.80 --72 33 55.2&  19 32 48.77 --72 33 53.3 & ELG  & 2 &0.287 & 18.8 & ESO05/02   \\
XBSJ204043.4--004548 & 20 40 43.46 --00 45 48.2&  20 40 43.46 --00 45 51.9 & NEL AGN   & 2 &0.615 & 21.2$^*$  &  ESO10/02   \\
XBSJ205635.7--044717 & 20 56 35.75 --04 47 17.9&  20 56 35.67 --04 47 17.2 & BEL AGN   & 1 &0.217 & 17.3 & TNG09/02   \\
XBSJ205829.9--423634 & 20 58 29.97 --42 36 35.0&  20 58 29.89 --42 36 34.2 & NLSy1  & 1 &0.232 & 18.3 &  ESO10/02   \\
XBSJ213002.3--153414 & 21 30 02.32 --15 34 14.1&  21 30 02.30 --15 34 13.2 & BEL AGN  & 1 &0.562 & 17.3 &  ESO10/02   \\
XBSJ213824.0--423019 & 21 38 24.03 --42 30 19.2&  21 38 24.03 --42 30 17.5 & BEL AGN   & 1 &0.257 & 16.6 &  (f)   \\
XBSJ214041.4--234720 & 21 40 41.45 --23 47 20.1&  21 40 41.53 --23 47 19.3 & BEL AGN   & 1 &0.490 & 18.4 & ESO10/02   \\
XBSJ220601.5--015346 & 22 06 01.50 --01 53 46.9&  22 06 01.45 --01 53 46.1 & BEL AGN   & 1 &0.211 & 20.1 & ESO10/02   \\
            \noalign{\smallskip}
            \hline
\end{tabular}

column 1: name;

column 2: X-ray position;

column 3: Optical position;

column 4: spectroscopic classification (BEL AGN=Broad Emission Line AGN; NEL AGN=Narrow Emission Line 
AGN; NLSy1=Narrow Line Seyfert 1; Sy1.9=Seyfert 1.9; ELG=Emission Line Galaxy);

column 5: optical spectral type;

column 6: redshift;

column 7: APM red magnitude; the asterisk indicates that the magnitude (R)
has been derived from our own observations;

column 8: reference for the identification i.e. the observing run, if the
source has been identified by us, or a reference, if the source 
has been identified from the literature. The notes are:

(a) V* BL Hyi. See Appendix A.1;

(b) identification from Severgnini et al. (2003)

(c) identification from La Franca, Cristiani \& Barbieri (1992)

(d) identification from Fiore et al. (2000)

(e) identification from Carter \& Malin (1983)

(f) identification from Hewett, Foltz \& Chaffee (1995)

AXIS: identification from the AXIS project (Barcons et al. 2002a,b)

\end{table*}

   \begin{table*}
      \caption[]{Spectroscopic classification}
         \label{id}
\begin{tabular}{l l r r}
   \noalign{\smallskip}
           \hline
class & Spectral type & Number & \% extragalactic \\ 
 \noalign{\smallskip}
           \hline\hline
{\bf Type~1}    &   & &  \\
\hline
          &  Broad Emission Line AGN  & 17  & 63\%\\
          &  NLSy1           &  2  & 7\% \\
\hline
          &   & {\bf 19} & {\bf 70\%} \\
\hline\hline
{\bf Type~2}    &  &  &  \\
\hline
          &  Narrow Emission Line AGN      &  5  & 19\% \\
          &  ELG            &  1  & 4\% \\
          &  Seyfert 1.9 &  2  & 7\% \\
\hline
          &  & {\bf 8} & {\bf 30\%} \\
\hline\hline
{\bf Accreting binary} &     & {\bf 1}  & \\
\hline            
	    {\bf Total}    &        &{\bf  28}   \\
            \noalign{\smallskip}
            \hline
         \end{tabular}
   \end{table*}

\section{X-ray data}

\begin{figure}
   \centering
   \includegraphics[width=9cm]{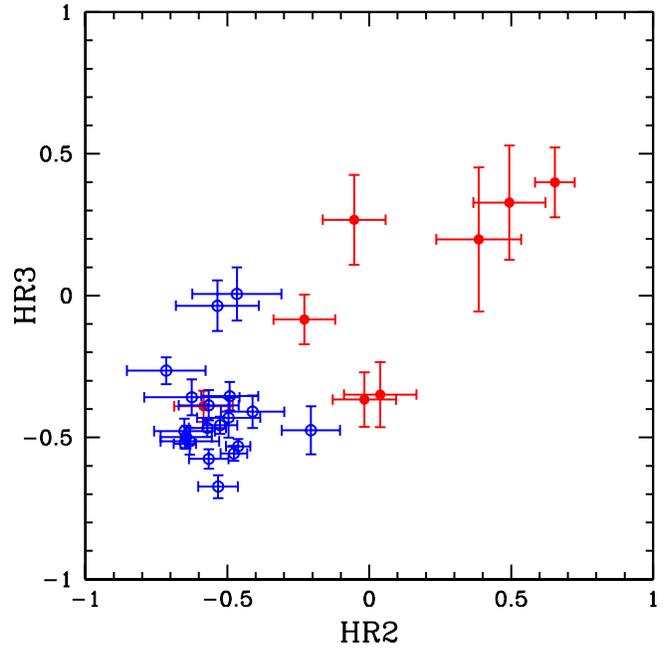}
      \caption{The hardness-ratio HR3 versus HR2. Filled circles represent the 
type~2 objects while the 
open circles are the type~1 objects
              }
         \label{hr2hr3}
   \end{figure}

On the basis of the MOS2 data, used for the sample definition, we have 
firstly studied the hardness ratios of the sources in the sample. 
This is useful as reference for other
(deeper) samples for which the available number of counts per source
is not enough to perform a detailed spectral analysis. 
The hardness ratios have been computed using the data in the 0.5-2.0 keV, 
2.0-4.5 keV and 4.5-7.5 keV
energy intervals adopting the following standard definitions:

\begin{center}

HR2 =$ \frac{C(2.0-4.5) - C(0.5-2.0)}{C(2.0-4.5) + C(0.5-2.0)}$

\end{center}
\begin{center}

HR3 = $\frac{C(4.5-7.5) - C(2.0-4.5)}{C(4.5-7.5) + C(2.0-4.5)}$

\end{center}

where C represents the  MOS2 corrected count-rates (see http://xmmssc-www.star.le.ac.uk/ 
for details) in the energy
intervals indicated in parenthesis (in keV). The hardness ratios 
used are weakly dependent on the Galactic N$_H$ (which is below 10$^{21}$ cm$^{-2}$
in the fields used for the sample selection). In Figure~\ref{hr2hr3} the 
HR3 is plotted against HR2. The figure clearly shows that the
optical classification and the hardness-ratio (HR2 in particular) 
are coupled: nearly all (7/8) the type~2 
objects show HR2 $>$--0.3 while  all but one type~1 sources 
have HR2 values below this value. The two exceptions are: XBSJ031146.1$-$550702,
a Sy1.9 with HR2=--0.58 and XBSJ031859.2$-$441627, a 
type~1 AGN with HR2=--0.21. As discussed below, the
X-ray analysis has confirmed the absence of a significant 
absorption in the first source and the presence of a mild 
absorption in the second one.  

Qualitatively, the correlation between the hardness-ratio and the optical spectral type 
is expected in the context of the unified models. A more
quantitative assessment of this correlation is done in the following Section 
on the basis of a detailed spectral analysis.

\subsection{X-ray spectral analysis}

Thanks to the relatively bright flux limit which characterizes
the sample, most of the sources have enough counts 
to perform a reliable X-ray spectral analysis. Except for 2 sources,
the total counts in the MOS2 range from 100 to 3900. 
The availability of the MOS1 and, in most cases, of the  pn
increases the counts by a factor $\sim$2 (if only the MOS1 is available) or $\sim$4 
(if both MOS1 and  pn are available) the actual number of counts 
available for the spectral analysis. For 7 sources 
the  pn was either not available or the object was outside the
field of view or under a bad column. 
In another case the source
falls under a gap in the MOS1 camera. In Table~\ref{xraydata}
the EPIC detectors used for the X-ray analysis of each source 
are summarized. Before the extraction of the spectra, 
we have removed the time intervals characterized 
by a high background. 

All the spectra have been extracted using a circular region with
a radius of 20$\arcsec$-30$\arcsec$, depending on the source off axis 
distance. The background
is extracted in a nearby source-free circular region of a radius 
a factor $\sim$2 larger than the one used to extract the source counts.
When the same blocking filters are used 
in the two MOS cameras the data are combined after the
extraction. 

The MOS and  pn data have been re-binned in order to 
have at least 15-25 counts per channel depending on the brightness
of the source. XSPEC 11.2 was then used to fit simultaneously the MOS and 
 pn data, leaving the relative normalization free to vary. 
During the fitting procedure only the data in the 0.3-10 keV and 0.4-10 keV energy ranges 
are considered for the MOS and the  pn respectively. 
Errors are given at the 90\% confidence level for one interesting
parameter ($\Delta\chi^2$=2.71).

Initially, we attempted to fit a simple absorbed power-law model to  
all the 27 spectra taking into account both the Galactic hydrogen column 
density along the line of sight (from Dickey \& Lockman 1990) and 
a possible intrinsic absorption at the source redshift. 

For 20 sources
the resulting fit is good enough (i.e. no systematic 
trends are observed in the residuals) and kept as the best-fit model
while for the remaining 7 sources a simple  absorbed power-law model does
not give an acceptable fit.

In most (13) of the 20 objects that are well fitted by a simple absorbed power-law model, 
no intrinsic absorption is found while for 7 sources the best-fit intrinsic N$_H$ is 
larger than the local Galactic column density. Among these latter sources, 
4 present an Hydrogen column  density larger than 10$^{22}$ cm$^{-2}$ 
and the related best fit photon index is flatter
($\Gamma<$1.5) than the flattest $\Gamma$ found for the unabsorbed AGNs (see below).  
This is probably due to the poor statistics available for spectral
analysis which do not allow us to adequately constrain at the same time both the 
spectral index and the absorption. 
In order to better constrain the correct value of N$_H$ we have 
fixed for these 4 sources the $\Gamma$ to 1.9 
(which corresponds to the average value found for the 
type~1 AGNs, see below) and derived the corresponding column density N$_H$. This procedure
is justified in the context of the unified scheme according to which the absorbed
AGNs should have the same intrinsic power-law and a larger absorption with respect
to the type~1 AGNs.
Since the values of $\Gamma$ found for the type~1 AGNs have an intrinsic
dispersion, we have checked the variation of N$_H$ when extreme values of
$\Gamma$  (within the observed distribution) are assumed. We have found
that the variation of N$_H$ is usually within the reported errors. 

Finally, for the 7 sources for which the simple absorbed power-law model
does not offer a good representation of the data, an improved fit is obtained 
with a more complex model, like a
``leaky absorbed power-law'' model\footnote{this model consists of two 
power-laws, one absorbed and one unabsorbed, having the same photon index}
or by adding a thermal component. In one case (XBSJ040758.9--712833) the
value of $\Gamma$ in the ``leaky absorbed power-law'' 
has been fixed to 1.9 to better constrain the absorption.

The results of the X-ray spectral analysis for the extragalactic sources 
are summarized in Table~\ref{xraymodel}. The X-ray spectra of the 
extragalactic sources are reported in Appendix \ref{xspec}.

\begin{figure}
   \centering
   \includegraphics[width=9cm]{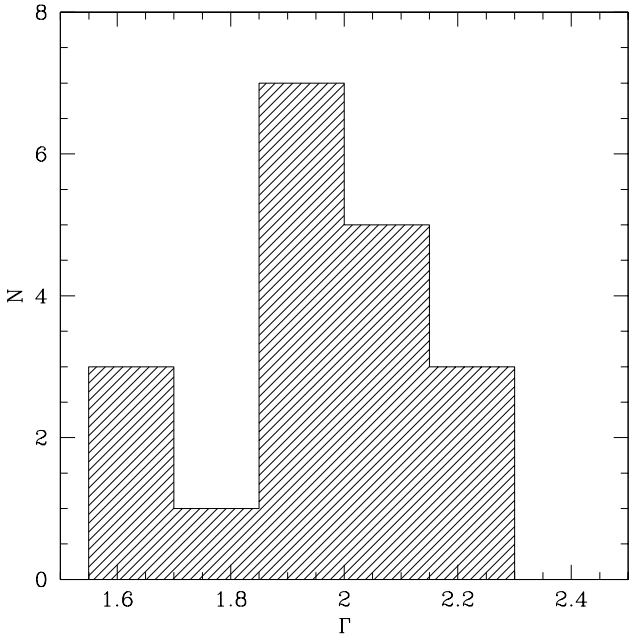}
      \caption{The X-ray photon-index distribution of the type~1 objects 
obtained from the spectral fitting.  
              }
         \label{gamma}
   \end{figure}

The distribution of the photon indices of the type~1
objects is shown in Figure~\ref{gamma}. The mean value is $<\Gamma>$=1.97
with a standard deviation of 0.16 consistent with the mean value found
in other similar samples (e.g. Mainieri et al. 2002).
The distribution 
of the photon indices of the type~2 objects is not presented since, 
for most of these sources, the value of $\Gamma$  has been fixed to 
the value 1.9 to better constrain the absorption.

We discuss below in greater detail the results for the single classes 
of extragalactic objects.

\subsection{Type 1 AGN}

{\bf Broad Emission Line AGNs.}
For 12 out of 17 Broad Line AGNs the simple absorbed power-law model
gives an acceptable fit (reduced $\chi^{2}$ below 1.2) and a column 
density consistent with the Galactic value. In one object (XBSJ031859.2--441627)
an intrinsic column density (N$_H$=3.9$\times$10$^{21}$ cm$^{-2}$ with $\Gamma$=1.72) 
significantly larger than the Galactic value results from the fitting 
procedure. This is one of two sources mentioned at the beginning of 
Section~4 for which  a departure from the HR2/optical 
classification segregation has been noted. 
The optical spectrum of this source is dominated by the
host galaxy while the AGN activity is suggested only by the presence of a
weak (EW$\sim$16 \AA), possibly broad H$\alpha$ emission line. 
In Severgnini et al. (2003) we show that the observed optical 
spectrum of this object can be modeled by using the N$_H$ computed from
the X-ray data and assuming a standard Galactic value for the E$_{B-V}$/N$_H$ ratio. 

In three other objects (XBSJ021808.3-045845, 
XBSJ205635.7-044717, XBSJ220601.5-015346) the simple absorbed
power-law model leaves significant residuals. A better fit is 
obtained by adding a soft component (a Black-body 
spectrum with a temperature of 0.1-0.2 keV). 

In summary, all the objects with broad (FWHM$>$2000 km s$^{-1}$)
emission lines in the optical spectra  are unabsorbed or only 
weakly absorbed in the X-ray band (N$_H<$4$\times$10$^{21}$ cm$^{2}$). 
In 18\% of cases, a soft excess is also detected.

{\bf Narrow Line Seyfert 1.}
The 2 NLSy1 candidates found in the sample  show only a moderate intrinsic absorption 
(N$_H$=9.5$\times$10$^{20}$ cm$^{-2}$, XBSJ205829.9-423634) or no
absorption at all (XBSJ023713.5-522734). In the case of XBSJ023713.5-522734,
a better fit is obtained by adding a soft-component modeled with a black-body 
spectrum with kT=0.1 keV. The spectral indices in the two sources
are quite similar ($\Gamma$=1.90$^{+0.18}_{-0.13}$ for XBSJ023713.5-522734 
and  $\Gamma$=1.91$^{+0.09}_{-0.09}$ for XBSJ205829.9-423634) and slightly
lower but still consistent with those found with ASCA by Vaughan et al. 
(1999) for a sample of 22 NLSy1 ($\Gamma$=2.1 with a standard deviation of 0.3 
in the 2-10 keV band). 

In summary, the 2 NLSy1 candidates are not significantly 
absorbed in the X-ray (N$_H<$10$^{21}$ cm$^{-2}$)
and, in one case, a soft-excess is also present. With the
available data there is no evidence that NLSy1 have steeper $\Gamma$ values 
when compared to Sy1.

\subsection{Type2 AGN}

{\bf Narrow Line AGNs and ELG.}
The X-ray spectra of all the Narrow Line AGNs and the ELG show strong 
evidence of intrinsic absorption (N$_H>$10$^{22}$ cm$^{-2}$). 

In two cases, a more complex model, including a ``scattered'' 
component (``leaky absorbed power-law'' model) is needed 
to obtain a better fit to the data. The scattered
component is found to be 5\% --10\% of the flux of the primary component
in agreement with what has been observed in other well studied Type~2 AGNs 
(e.g. Turner et al. 1997a,b).

Finally, in one case (XBSJ033845.7-352253) the leaky absorbed power-law
model does not give  an acceptable fit leaving significant residuals
at low energies. The inclusion of a thermal component with 
kT=0.16$^{+0.04}_{-0.08}$ keV is required to achieve a better fit (see
Appendix~A.3 for a more detailed discussion of this source).

In summary, the 6 objects with an optical spectrum showing
only narrow emission lines (FWHM$<$1000 km s$^{-1}$) are significantly
absorbed  at X-ray energies (N$_H>$10$^{22}$ cm$^{2}$).

\begin{figure}
   \centering
   \includegraphics[width=9cm]{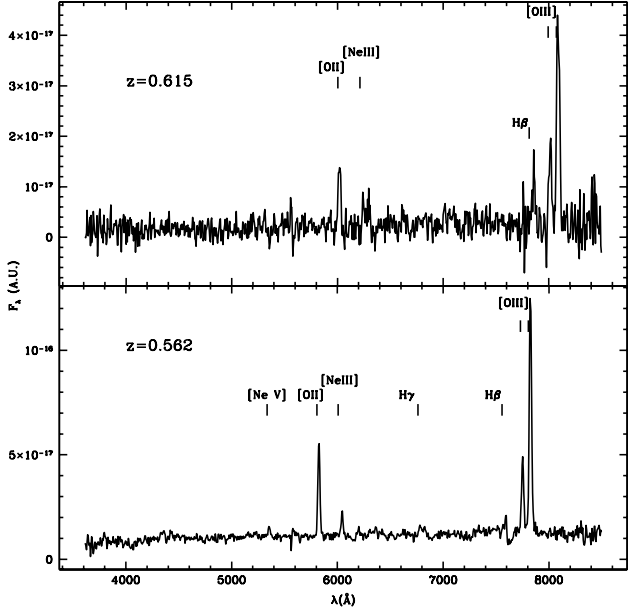}
      \caption{The optical spectra of the 2 type~2 QSO discovered in the 
HBS28 sample: bottom panel: XBSJ013240.1-133307, top panel: XBSJ204043.4-004548
              }
         \label{qso2}
   \end{figure}

Notably, two  objects (XBSJ013240.1-133307 and XBSJ204043.4-004548)
have an unabsorbed X-ray luminosity (L$_{2-10 keV}$=2.6$\times$10$^{44}$ erg s$^{-1}$ 
and L$_{2-10 keV}$=3.2$\times$10$^{44}$ erg s$^{-1}$ respectively) in 
the typical range of the type~2 QSO. The optical spectra of these
two sources are shown in Figure~\ref{qso2}.

%

{\bf Seyfert 1.9.}
The 2 ``intermediate'' Seyfert galaxies show  very different 
spectra:  XBSJ040658.8-712457 is absorbed 
($N_H$=2.5$\times$10$^{23}$ cm$^{-2}$) while XBSJ031146.1-550702 
is well fitted by a simple un-absorbed power-law model
with $N_H<$1.3$\times$10$^{20}$ cm$^{-2}$. The very low  X-ray 
absorption  observed in this latter source is
in contrast with what is usually found in Sy1.9 for which
typical values of $N_H$ are between 10$^{22}$ and 
10$^{24}$ cm$^{-2}$ (e.g. Risaliti, Maiolino \& Salvati 1999; 
Maiolino et al. 2001). The source XBSJ031146.1-550702 seems
to be a case where the measured X-ray absorption is much lower than 
the one inferred by the optical spectrum, which shows no broad lines
except for a broad wing in the H$\alpha$ profile and an 
[OIII]$\lambda$5007\AA/H$\beta$ ratio typical of a Sy2. 
Other cases like this have been presented in the literature (see next
section). 
A more detailed discussion of this object is reported in
Appendix~A.2.

   \begin{figure}
   \centering
   \includegraphics[width=9cm]{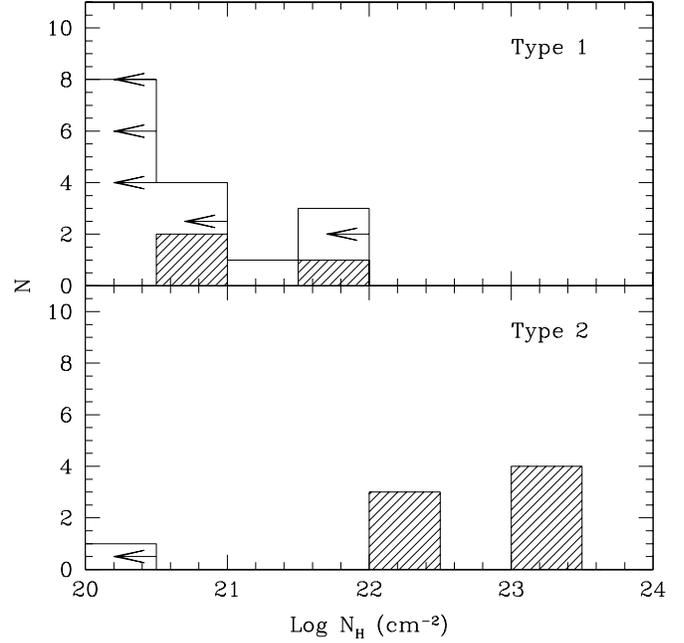}
      \caption{The distribution of the intrinsic N$_H$ derived from the
spectral fit, for Type~1 objects 
and Type~2 objects.
The shaded histograms show the detections while the empty
histograms represent the upper limits on N$_H$.
              }
         \label{nh}
   \end{figure}
%

   \begin{figure*}
   \centering
   \includegraphics[width=13cm,angle=-90]{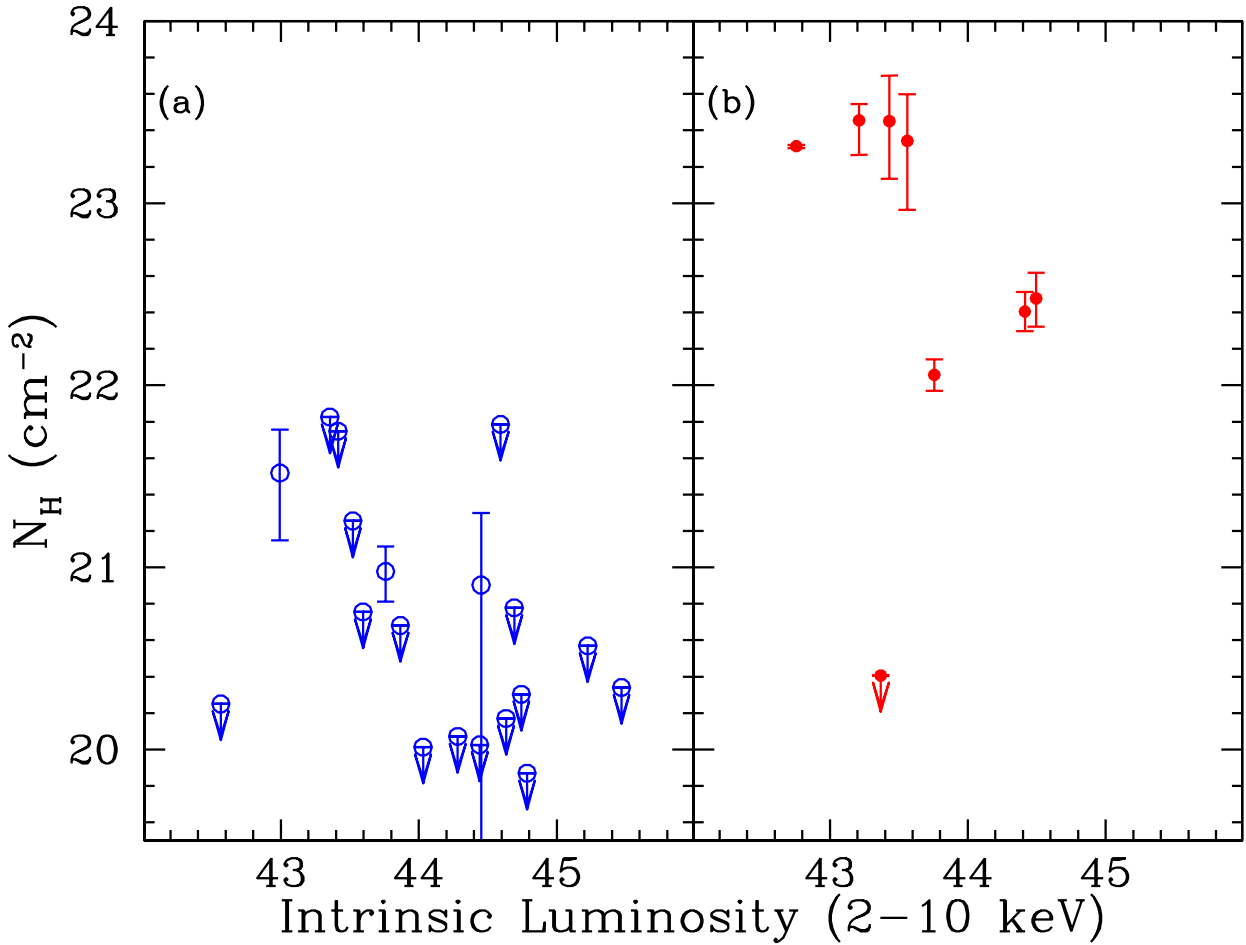}
      \caption{The intrinsic N$_H$ versus the unabsorbed X-ray
               luminosity in the 2-10 keV energy range 
              for (a) Type~1 objects (open circles) and
              for (b) the Type~2 objects (filled circles). 
              }
         \label{lx_abs}
   \end{figure*}


\begin{table*}
\caption{Best-fit parameters for the X-ray spectral analysis (extragalactic objects)}
\label{xraymodel}
\begin{tabular}{l c l r r l l r  r }
          \noalign{\smallskip}
          \hline
name & id & Model & $\Gamma$ & N$_H$   & kT         & red $\chi^2$/dof & f$_{2-10 keV}$ & Log L$_{2-10 keV}$\\ 
     &    &   &          & ($\times$10$^{22}$) &      &         &($\times$10$^{-13}$) & \\
            \noalign{\smallskip}
            \hline
XBSJ002618.5$+$105019 & BEL AGN & PL &  2.02$^{+ 0.07}_{-0.06}$ & $<$  0.011 & ... & 1.15 /  97 &   2.8 & 44.44 \\
XBSJ013240.1$-$133307 & NEL AGN &PL & 1.9 (frozen) &    2.54$^{+   0.71}_{  -0.56}$ & ... & 1.11 /  19 &   1.7 & 44.41 \\
XBSJ013944.0$-$674909 & BEL AGN &PL &  1.94$^{+ 0.14}_{-0.12}$ & $<$  0.018 & ... & 0.91 /  34 &   1.2 &  42.56 \\
XBSJ015957.5$+$003309 & BEL AGN &PL &  2.14$^{+ 0.09}_{-0.09}$ & $<$  0.010 & ... & 1.19 /  99 &   2.9 &  44.03 \\
XBSJ021640.7$-$044404 & BEL AGN &PL &  2.24$^{+ 0.08}_{-0.09}$ & $<$  0.020 & ... & 1.17 /  67 &   1.1 & 44.74 \\
XBSJ021808.3$-$045845 & BEL AGN &PL + BB &  2.01$^{+ 0.04}_{-0.04}$ & $<$  0.007 &   0.16$^{+  0.01}_{ -0.01}$ & 1.07 / 527 &   2.3 & 44.79 \\
XBSJ021817.4$-$045113 & BEL AGN &PL &  1.84$^{+ 0.04}_{-0.03}$ & $<$0.04 & ... & 1.07 / 346 &   2.7 &  45.22 \\
XBSJ021822.2$-$050615$^a$ & NEL AGN &Leaky &  1.66$^{+ 0.34}_{-0.36}$ &   20.54$^{+   0.36}_{  -0.44}$ & ... & 1.41 /  63 &  3.0 & 42.76 \\
XBSJ023713.5$-$522734 & NLSy1 &PL + BB &  1.90$^{+ 0.18}_{-0.13}$ & $<$  0.18 &   0.11$^{+  0.02}_{ -0.02}$ & 0.86 /  80 &   2.7 & 43.52  \\
XBSJ030206.8$-$000121 & BEL AGN &PL &  1.90$^{+ 0.06}_{-0.06}$ & $<$  0.015 & ... & 1.14 / 125 &   2.2 & 44.63\\
XBSJ030614.1$-$284019$^b$ & BEL AGN &PL &  1.56$^{+ 0.26}_{-0.13}$ & $<$  0.048 & ... & 1.15 /  16 &   2.9 & 43.87 \\
XBSJ031015.5$-$765131 & BEL AGN &PL &  1.92$^{+ 0.04}_{-0.04}$ & $<$  0.022 & ... & 0.90 / 189 &   3.4 & 45.47 \\
XBSJ031146.1$-$550702 & Sy1.9 &PL &  2.08$^{+ 0.13}_{-0.12}$ & $<$  0.013 & ... & 0.93 /  44 &   2.8 &  43.37 \\
XBSJ031859.2$-$441627$^a$ &BEL AGN & PL & 1.58$^{+ 0.29}_{-0.29}$ & 0.33$^{+0.43}_{-0.19}$ & ... & 0.73 /  18 &   1.7 &  42.99 \\
XBSJ033845.7$-$352253 & NEL AGN &Leaky + RS & 1.9 (frozen) &   28.5$^{+   6.7}_{-10.0}$ &   0.16$^{+  0.04}_{ -0.08}$ & 0.83 /  29 &   1.7 &  43.21 \\
XBSJ040658.8$-$712457 & Sy1.9 & PL & 1.9 (frozen) &   21.97$^{+  17.67}_{  -12.78}$ & ... & 1.25 /   4 &   1.6 &  43.56 \\
XBSJ040758.9$-$712833 & NEL AGN &Leaky &  1.9 (frozen) &   28.23$^{+21.96}_{-14.64}$ & ... & 1.13 /   5 &   2.5 & 43.44 \\
XBSJ041108.1$-$711341 & BEL AGN &PL &  1.95$^{+ 0.48}_{-0.36}$ & $<$  0.61 & ... & 0.99 /   6 &   0.8 & 44.59 \\
XBSJ185613.7$-$462239 & BEL AGN &PL &  2.18$^{+ 0.27}_{-0.24}$ & $<$  0.058 & ... & 1.15 /  26 &   1.4 &  44.69 \\
XBSJ193248.8$-$723355 & ELG & PL & 1.9 (frozen) &    1.14$^{+   0.25}_{  -0.21}$ & ... & 1.06 /  35 &   1.9 &  43.76 \\
XBSJ204043.4$-$004548 & NEL AGN &PL & 1.9 (frozen) &    3.00$^{+   1.15}_{  -0.90}$ & ... & 1.35 /  16 &   1.7 & 44.50 \\
XBSJ205635.7$-$044717 & BEL AGN &PL + BB &  1.91$^{+ 0.51}_{-0.35}$ & $<$  0.56 &   0.15$^{+  0.03}_{ -0.03}$ & 1.18 /  21 &   1.7 &  43.42 \\
XBSJ205829.9$-$423634 & NLSy1 &PL &  1.91$^{+ 0.09}_{-0.09}$ &    0.09$^{+   0.04}_{  -0.03}$ & ... & 0.76 / 120 &   3.2 &  43.76 \\
XBSJ213002.3$-$153414 & BEL AGN &PL &  2.10$^{+ 0.23}_{-0.21}$ &    0.08$^{+   0.12}_{  -0.08}$ & ... & 0.82 /  26 &   1.8 &  44.45 \\
XBSJ213824.0$-$423019$^c$ & BEL AGN &PL &  2.14$^{+ 0.25}_{-0.17}$ & $<$  0.057 & ... & 0.86 /  26 &   1.6 & 43.60 \\
XBSJ214041.4$-$234720 & BEL AGN &PL &  2.19$^{+ 0.10}_{-0.10}$ & $<$  0.012 & ... & 1.06 /  93 &   1.7 & 44.28 \\
XBSJ220601.5$-$015346 &BEL AGN & PL + BB &  1.56$^{+ 0.16}_{-0.09}$ & $<$0.670 & 0.06$^{+  0.01}_{ -0.04}$ & 0.72 /  15 &   1.7 & 43.36 \\
            \noalign{\smallskip}
            \hline
\end{tabular}

column 1: source name;

column 2: optical identification;

column 3: best-fit model (PL = absorbed power-law; Leaky = leaky absorbed power-law;
PL + BB = absorbed power-law plus Black-Body spectrum; Leaky + RS = leaky absorbed power-law plus a Raymond-Smith
spectrum)

column 4: best-fit photon index of the power-law component;

column 5: best-fit intrinsic hydrogen column density [cm$^{-2}$];

column 6: temperature of the thermal component (if present) [keV];

column 7: reduced Chi-squared and degrees of freedom of the best-fit;

column 8: observed 2-10 keV flux (corrected for the Galactic absorption) [erg s$^{-1}$ cm$^{-1}$];

column 9: Logarithm of the intrinsic luminosity in the 2-10 keV energy range [erg s$^{-1}$];

$^a$ results from the X-ray analysis taken from Severgnini et al. (2003)

$^b$ in this object the presence of an emission line at 1.9 keV 
(source rest frame, consistent with the Si XIII at 1.86 keV) is required.

$^c$ in this object the presence of an emission line at 6.4 keV 
(source rest frame, consistent with the Fe I K$\alpha$) 
is required. 
\end{table*}


\section{Optical versus X-ray absorption}

The unified models (Antonucci 1993) predict a strict correlation 
between the optical and the X-ray
absorption. Although the presence of this correlation is 
suggested by the large values of N$_H$ usually observed in Seyfert 2 
galaxies (e.g. Awaki et al. 1991; Turner et al. 1997a; 
Risaliti et al. 1999; Guainazzi et al. 2001), 
many important exceptions are known since the first observations 
taken with the {\it Einstein} Observatory (e.g. Maccacaro, Perola \& Elvis 1982).
More recently, Panessa \& Bassani (2002)
have presented a significant number of type~2 AGNs whose X-ray
spectra are indicative of a very low (or even absent) absorption 
(N$_H<$10$^{22}$ cm$^{-2}$), thus confirming earlier results 
from Ptak et al. (1996), Pappa et al. (2001) and Bassani et al. (1999).
Similarly, Barcons, Carrera \& Ceballos (2003) have recently shown
the case of a Sy1.9 with no X-ray absorption and more cases like 
this are emerging from recent XMM observations (Lehmann et al. in prep.).
Quantifying the percentage of this type of source is
a difficult task. However, Panessa \& Bassani (2002)
estimate this percentage in the range 10\%-30\%. This number
is  larger than that found by Risaliti et al. (1999)
in a sample of Seyfert~2 (4\%). 

Another important exception to the optical/X-ray absorption 
correlation is the discovery of type~1 AGNs with large X-ray
absorption (e.g. Fiore et al. 2001). 
Similarly, Maiolino et al. (2001) have pointed out that in a fraction
of AGNs the $A_V/N_H$ ratio is significantly lower than
that observed in our Galaxy. 

The statistical completeness of the HBS28 sample together with
its complete spectroscopic classification allows us to perform
a statistical analysis of the relationship between X-ray and
optical absorption at the flux limit of the survey  
($\sim$4-7$\times$10$^{-14}$ erg s$^{-1}$ cm$^{-2}$ in the
4.5-7.5 keV energy range). 
In Figure~\ref{nh} the distribution of the intrinsic N$_H$ for the extragalactic 
objects  is presented. The number of X-ray obscured AGNs (N$_H>$10$^{22}$ cm$^{-2}$) is 7 
corresponding to a surface density of 0.7 deg$^{-2}$. 

In Figure~\ref{lx_abs} the best-fit values of 
N$_H$  are plotted against the 2-10~keV intrinsic luminosity.
In panel (a) of Figure~\ref{lx_abs} only the sources 
optically classified as type~1 objects are plotted while
in panel (b) only the type~2 objects are presented. 
The plot confirms what has been
already inferred from the analysis of the hardness ratios, i.e. that
a correlation between the optical type and the X-ray absorption is 
clearly present: type~1 objects always have  low values of 
N$_H$ ($<$10$^{22}$ cm$^{-2}$,  panel a) while the majority (88\%) of type~2
objects have large values of N$_H$ ($>$10$^{22}$ cm$^{-2}$) (panel b). 
This correlation holds also in the other sense: the great majority (95\%) of sources 
with N$_H<$10$^{22}$ cm$^{-2}$ are type~1 while all 
the objects with N$_H>$10$^{22}$ cm$^{-2}$ are type~2 sources.

In the HBS28 sample one source out of eight type~2 sources has
N$_H<$10$^{22}$ cm$^{-2}$ (see previous Section). Hence, we can infer that 
the fraction of these sources is 12\% with a standard deviation range of 
2\% - 41\%. 
Given the large uncertainity, this result is consistent with both 
the estimate made by Panessa \& Bassani (2002) and with that
of Risaliti et al. (1999). 

Moreover, we have found two type~1 AGNs 
in which the absorption exceeds the Galactic 
column density, but in only one of these the intrinsic absorption 
is relatively large (N$_H$=3.3$\times$10$^{21}$ cm$^{-2}$, XBSJ031859.2--441627). 
As already discussed, also in this object the observed 
N$_H$ and the optical spectrum are consistent with a standard Galactic $A_V$/$N_H$ 
(Severgnini et al. 2003). 

Interestingly, we have not found any type~1 object for which the value of N$_H$ is higher
than 10$^{22}$ cm$^{-2}$ (see Figure~\ref{nh}).
Based on the available numbers we can estimate an upper limit of 10\% on
the percentage of type~1 sources with N$_H>$10$^{22}$ cm$^{-2}$. 

\section{Comparison with the XRB  synthesis models}
The relative percentage of absorbed versus un-absorbed AGNs found
in the HBS28 sample can be directly compared to the  theoretical  predictions of 
the XRB  synthesis models. 
We stress once again the great advantage of 
dealing with a completely identified sample for which a value for 
the redshift is available for all sources. This is fundamental 
for a correct determination of the intrinsic absorbing column
density. 

The 4.5-7.5 keV selection interval is useful for probing the real fraction of 
obscured objects even at relatively bright flux limits. 
The expected fraction of obscured AGNs 
(N$_H>$10$^{22}$ cm$^{-2}$)
in the HBS28  predicted by the XRB synthesis model discussed in Comastri et 
al. (2001) is equal to 0.62-0.65, considering the range of fluxes 
sampled by the HBS28 (between 0.4 to 2.5$\times$10$^{-13}$ erg s$^{-1}$ cm$^{-2}$). 
This fraction corresponds to 17-18 
AGNs obscured in the X-rays expected in the HBS28 sample. 
This number is significantly (at the 99\%
confidence level) larger than the observed one (7), assuming the Poissonian statistics 
(Gerhels 1986). 

A similar overprediction, although less significant (90\%), of the population 
of absorbed AGNs is also found by comparing our results with the ``modified unified model'' recently 
presented by Ueda et al. (2003), which includes a dependence of the fraction of
absorbed AGNs on the X-ray luminosity. 

In the 2-10 keV energy band Piconcelli et al. (2002, 2003) have already 
pointed out that the observed fraction of the absorbed AGNs at a flux limit
of $\sim$10$^{-14}$ erg s$^{-1}$ cm$^{-2}$ is a factor $\sim$2 lower that 
predicted by the CXB synthesis models.
We thus confirm this result and extend it to the hardest  4.5-7.5 keV
energy band. 
Clearly, current XRB synthesis models have to be revised.

\section{Type 2 QSO}
\begin{figure}
   \centering
   \includegraphics[width=9cm]{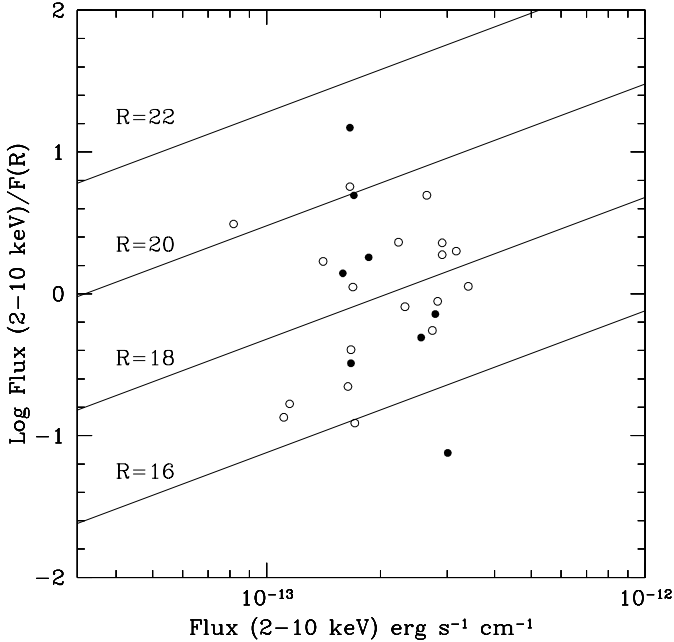}
      \caption{The ratio between the (observed) flux between 2 and 10 keV
and the optical flux (see text for details) versus the observed 2-10 keV flux. 
The loci of constant R (Cousins) magnitudes are also plotted for R=16,18, 20 and 22. 
Symbols are the same as in  previous figure.
              }
         \label{x_o}
   \end{figure}

As already mentioned, two type~2 QSO have been discovered in the
HBS28 sample, namely XBSJ013240.1-133307 (z=0.562, intrinsic 
L$_{2-10 keV}$=2.6$\times$10$^{44}$ erg s$^{-1}$) 
and XBSJ204043.4-004548 (z=0.615, intrinsic L$_{2-10 keV}$=3.2$\times$10$^{44}$ erg s$^{-1}$).
In Figure~\ref{x_o} the ratio  between the observed flux 
(2-10 keV) and the optical flux\footnote{
F$_{opt}$=$\Delta\lambda\times$f$_{6400\AA}$ where f$_{6400\AA}$ is the 
monochromatic flux at 6400\AA\ derived from the APM red magnitudes
and $\Delta\lambda$ is the band width of the R Cousins filter. 
}  are plotted versus the observed 2-10 keV flux. The two 
type~2 QSO are among the three objects with the largest X-ray-to-optical
flux ratio (1.2 and 0.8) in the sample.
The surface density of type~2 QSO at the flux limit of the HBS28 
is 0.21$^{+0.15}_{-0.13}$ deg$^{-2}$.

%


\section{X-ray bright Optically Normal Galaxies (XBONG)}

The existence of optically ``passive'' galaxies for which the
X-ray observations reveal  an intense AGN activity has been
hotly discussed in the recent literature (e.g. Fiore et al. 2000;
Barger et al. 2002, Comastri et al. 2002; Severgnini et al. 2003).
Even if these X-ray bright Optically Normal Galaxies (XBONG) seem
to emerge at faint X-ray fluxes ($<$10$^{-13}$ erg cm$^{-2}$ s$^{-1}$),
some examples have been discovered also at relatively high X-ray fluxes
(e.g. Piconcelli et al. 2002; Severgnini et al. 2003). 

In the HBS28 sample two sources were at a first glance classified as 
XBONG candidates, based on
the optical information available at the time of the sample selection. 
The source XBSJ021822.2--050615
was initially classified as an XBONG since no strong emission 
lines were present in the discovery spectrum.
However, a more accurate re-observation of this source 
with the 8.2m Subaru telescope  has revealed 
the presence of narrow emission lines that were outshone by the host-galaxy
light in the previous spectrum (Severgnini et al. 2003). This
source is thus classified as Type~2 AGN.

Similarly, the source XBSJ031859--441627 was initially classified
as an ``optically dull galaxy'' on the basis of an optical spectrum covering 
the spectral range between $\sim$4900 and 5800\AA. Further observations,
covering a wider spectral range, have revealed  of 
$\Gamma$ the presence of a
weak and possibly broad H$\alpha$ emission line (Severgnini et al. 2003).
We consider this object to be a type~1 AGN.

Interestingly, the two objects have the lowest X-ray-to-optical flux ratio
(and the brightest magnitude) among the objects in the HBS28 sample 
(see Figure~\ref{x_o}). This peculiar X-ray-to-optical flux ratio  
found for this type of sources has been already noted, for instance,
by Comastri et al. (2003). 

In conclusion, two sources in the HBS28 sample are potentially
similar to some XBONG that are emerging from different X-ray surveys
although a closer look at their spectra, based on high resolution
data and/or a better spectral coverage, reveal their real AGN nature.
We stress that similar detailed observations, relatively easy to do for low-z sources (like those
discovered in the HBS28 sample), are hard to carry out for higher z objects.

\section{Summary and conclusions}

We have presented a statistically complete sample of 28 X-ray
sources selected in the 4.5-7.5 keV energy band at a flux limit
of $\sim$4-7$\times$10$^{-14}$ erg s$^{-1}$ cm$^{-2}$. 
The sample is completely identified and a spectroscopic
classification and redshift is available for each object.
This is the first sample, selected in this energy band, to 
have such complete optical information. Apart from one
Galactic accreting binary star, all the remaining 27 sources are AGNs. 
We have discussed the optical classification of these 27 objects
based on the presence and the intensities of the broad/narrow emission
lines. In particular, we have divided the sources into 19 type~1 objects, 
including Sy1, QSO and NLSy1, and 8 type~2 objects, including 
Sy2, Sy1.9, Narrow Line Radio Galaxies (NLRG) and ELG. 
We have then presented the X-ray spectral
analysis based on the XMM-{\it Newton}
 EPIC data  for the 27 AGNs. Thanks to the relatively bright flux limit of 
the sample, the available  number of counts (from hundreds to thousands) 
is large enough to allow a reliable spectral analysis in nearly all
cases.

The optical and the X-ray properties of the sources are thus analysed
together. The main results can be summarized as follows:

\begin{itemize}

\item A correlation between the optical spectral type and the
X-ray absorption is found: all the type~1 objects show column densities,
derived from the X-ray spectra, below 10$^{22}$ cm$^{-2}$. On the contrary, all
but 1 (88\%) of the type~2 objects are characterized by column densities
above 10$^{22}$ cm$^{-2}$. This result strongly supports the unified models
(e.g. Antonucci 1993);

\item Despite the mentioned good correlation, we have found one
type~2 object (optically classified as Sy1.9) with no absorption 
(N$_H<$1.3$\times$10$^{20}$ cm$^{-2}$) in the X-ray band. In
Appendix A.2 we have discussed the similarities of this object with those found 
by other authors (e.g. Panessa \& Bassani 2002);

\item Two type~2 QSOs, with an intrinsic luminosity larger than
10$^{44}$ erg s$^{-1}$ (2-10 keV) and N$_H>$10$^{22}$ cm$^{-2}$ 
have been discovered in the sample. These two objects represent  
25\%/30\%  of the optical/X-ray absorbed AGNs respectively. 
The corresponding surface density of type~2 QSO at the flux limit of 
the survey is 0.21$^{+0.15}_{-0.13}$ deg$^{-2}$;

\item The fraction of absorbed AGNs (N$_H>$10$^{22}$ cm$^{-2}$) in 
this sample (26\%) is significantly lower (at $\sim$90\% confidence level) 
than the fraction predicted by the current XRB synthesis models 
at this flux limit (Comastri et al. 2001; Ueda et al. 2003). The observed lack
of absorbed AGNs, recently suggested also by  other authors 
(e.g. Piconcelli et al. 2002, 2003), clearly indicates that the XRB
synthesis models should be revised, at least with respect to the 
relative fraction of obscured/un-obscured objects.

\end{itemize}

\begin{acknowledgements}
We thank Andrea Comastri for providing us with the predictions of the 
XRB synthesis model adapted to the 4.5-7.5 keV energy band. 
This work has received partial 
financial support  from ASI (I/R/062/02, I/R/037/01) 
and from the Italian Ministry of University and Scientific and Technological 
Research (MURST) through grant Cofin. XB and FJC acknowledge 
financial support from the Spanish Ministry of Science
and Technology, under project AYA2000-1690.
The TNG telescope is operated on the island of La Palma 
by the Centro Galileo Galilei of the INAF in the Spanish
Observatorio del Roque de Los Muchachos of the Instituto de Astrof\'\i
sica de Canarias. 
We would like to thank the staff members of the
ESO and TNG Telescopes for their support during the observations.
This research has made use of the NASA/IPAC Extragalactic Database 
(NED) which is operated by the Jet Propulsion Laboratory, California 
Institute of Technology, under contract with the National Aeronautics 
and Space Administration. We finally thank the APM team  
for maintaining this facility.
\end{acknowledgements}


\appendix
\section{Notes on individual objects}

\subsection{XBSJ014100.6--675328: The {\it polar} BL Hyi}

XBSJ014100.6--675328 (BL Hyi) is a known AM Herculis object 
(a {\it polar}), i.e. a binary system composed of a magnetic 
white dwarf and a low-mass star. These sources are characterized by soft and hard 
X-ray variable emission, modulated on the rotational period
of the white dwarf. The hard X-ray properties of BL Hyi
have been discussed by Wolff et al. (1999) on the basis 
of RXTE data and by Matt et al. (1998) by using ASCA and 
BeppoSAX observations. 

The X-ray spectrum of  XBSJ014100.6--675328 is well fitted
by a unabsorbed power-law model with $\Gamma$=1.53$^{+0.02}_{-0.03}$ 
plus a thermal component (kT=55$^{+3}_{-1}$ eV) and an emission
line at E=6.7$\pm$0.2 keV. The 2-10 keV flux is 
6.1$\times$10$^{-12}$ erg s$^{-1}$ cm$^{-2}$.

\subsection{XBSJ031146.1-550702}
XBSJ031146.1-550702 has been classified as Sy1.9 and its
X-ray emission does not show significant absorption, as described
in the text. 
Other examples of Sy2 objects with low X-ray absorption 
have been recently presented by Panessa \& Bassani (2002) 
and Barcons, Carrera \& Ceballos (2003).  
Panessa \& Bassani (2002) have suggested that  
a large-scale dusty environment (e.g. dust lanes or an HII region), 
instead of an obscuring torus close to the nucleus,
can hide the BLR without producing a significant X-ray absorption
(N$_H\sim$10$^{22}$ cm$^{-2}$ or slightly lower than this value).
However, the upper limit on the N$_H$ value estimated for 
XBSJ031146.1-550702 is 1.3$\times$10$^{20}$ cm$^{-2}$ which is
almost two orders of magnitude below the value expected in the 
scenario proposed by Panessa \& Bassani (2002). 
Another possibility considered by Panessa \& Bassani (2002) is
that the BLR is not efficiently activated due to an extremely
sub-Eddington accretion rate. This scenario, however, is expected 
to work for low-luminosity AGNs (LLAGN, L$_X<$10$^{42}$ erg s$^{-1}$) 
which represent a large fraction of the Sy2 studied by 
Panessa \& Bassani (2002). 
On the contrary, the source XBSJ031146.1-550702 has an X-ray
luminosity of 2.2$\times$10$^{43}$ erg s$^{-1}$, thus significantly
higher than the luminosity expected in LLAGNs. 
Two objects similar to XBSJ031146.1-550702 in terms of X-ray 
luminosity ($\sim$10$^{42}$-10$^{43}$ erg s$^{-1}$) 
and low N$_H$ value ($<$10$^{20}$- 10$^{21}$ cm$^{-2}$)  
are \object{NGC~7679} (Della Ceca et al. 2001)
and \object{IRAS00317-2142} (Georgantopoulos 2000). These two sources
are dominated by the starburst emission in the optical
and by an unabsorbed AGN  in the X-ray. Selective obscuration
due to a dusty ionized absorber has been invoked to explain the
observed X-ray and optical properties of these sources. 
A more detailed analysis of XBSJ031146.1-550702 is required to 
assess whether a similar interpretation could also be applied to 
this source. 

Another possibility is that XBSJ031146.1-550702 is a Compton-thick
source (N$_H>$10$^{24}$ cm$^{-2}$). 
In this case, the observed X-ray spectrum represent only the
fraction of the total emission reflected towards the observer. 
If this hypothesis is correct, a strong neutral iron K$\alpha$ line is expected at 6.4 keV
(source rest frame), corresponding to 5.5 keV in the observer 
rest frame, i.e. very close to the extreme end of the observed spectrum.
In this spectral region an upper limit on the emission line is very
poorly determined (EW$\leq$2~keV) and cannot be used to rule out
the Compton thick nature of the source. 

In principle, the
thickness parameter F$_X$/F$_{\rm{[OIII]}}$, where F$_X$ is the 2-10 keV 
observed flux (corrected for the Galactic absorption) and F$_{\rm{[OIII]}}$ 
is the reddening corrected flux of the [OIII]$\lambda$5007\AA\ line,
could be used to assess the Compton thickness of the source.
According to Bassani et al. (1999), in fact,  
a Compton-thick source should present low values of 
the F$_X$/F$_{\rm{[OIII]}}$ ratios (below 1) while Compton-thin 
sources should have F$_X$/F$_{\rm{[OIII]}}$ significantly above 1. 
As previously stated, the optical spectra collected 
by us do not have an absolute flux calibration so the
measurement of the F$_X$/F$_{\rm{[OIII]}}$  flux ratio is not
simple. However, considering the good weather conditions
of the night during the observation of XBSJ031146.1-550702 
and taking into account the flux lost due to the
slit width we have obtained a rough absolute calibration
for this spectrum. The result is in good agreement with the APM 
photometric point (red magnitude). We have thus estimated
the line flux and the  F$_X$/F$_{\rm{[OIII]}}$ flux ratio\footnote{
The value F$_{\rm{[OIII]}}$ should be corrected for the reddening 
based on the Balmer decrement as explained in Bassani et al. (1999).  
In XBSJ031146.1-550702, however, the estimate of the Balmer
decrement is difficult due to the presence of the broad H$\alpha$ 
component and the relatively poor S/N ratio of the available spectrum.
We attempted a de-blending of the two components in the H$\alpha$ 
profile and we have obtained H$\alpha$/H$\beta\sim$2--3 which
indicates that the reddening should not be very important. 
For this reason we have not applied any reddening correction. 
}. The computed ratio is between 8 and 30, hence suggesting a 
Compton-thin nature for this object. Clearly, better 
data are required to put this estimate on a firmer ground.

\subsection{XBSJ033845.7-352253}

The X-ray spectrum of  this NLRG is not fitted by a simple absorbed power-law 
or a leaky absorbed power-law models. A reasonable fit
is obtained by allowing the two spectral indices free to vary. 
The best fit gives a very steep slope ($\Gamma$ = 2.71) at low energies plus
a flatter slope ($\Gamma$ = 1.69) at higher energies. Alternatively,
a leaky absorbed (N$_H$=2.8$\times$10$^{23}$ cm$^{-2}$) power-law with $\Gamma$ frozen to 1.9 plus
a thermal (Raymond-Smith) model with kT=0.16$^{+0.04}_{-0.08}$ 
keV gives an acceptable fit. 
The simple absorbed power-law plus a thermal component, instead,
is unable to fit both the high and the low energy parts of the 
spectrum. 

XBSJ033845.7-352253 is a known Narrow Line Radio galaxy (NLRG,
Carter \& Malin 1983) resolved in the NRAO VLA Sky Survey 
(NVSS, Condon et al. 1998) into two bright radio-lobes with a total radio 
flux of 2.1 Jy corresponding to a radio 
power of 7.8$\times$10$^{25}$ W Hz$^{-1}$ at 1.4 GHz. 
In principle, the flat component ($\Gamma$ = 1.69) of the 2 power-laws model 
could be ascribed to the non-thermal emission from the jet. 
However, this component is usually
not observed in lobe-dominated radio-loud AGNs, like XBSJ033845.7-352253,
where relativistic beaming is not important while it is probably at the origin of the 
flat X-ray spectrum observed in
core-dominated AGNs (e.g. Sambruna, Eracleous \& Mushotzky 1999). 

We consider the second model, which includes a thermal emission, 
as more appropriate. In fact, the presence of a thermal emission is a quite frequent
feature  in NLRGs. For instance, 50\% of the NLRGs observed with ASCA  
(Sambruna, Eracleous \& Mushotzky 1999) show a thermal component;
the 0.1-2.4 keV unabsorbed luminosity of the thermal emission 
found in XBSJ033845.7-352253 is 2$\times$10$^{41}$ erg s$^{-1}$, 
in the range of luminosity
obtained by Sambruna, Eracleous \& Mushotzky (1999) (L$_X\sim$10$^{40-43}$ 
ergs s$^{-1}$). 

Therefore, we have adopted the thermal plus leaky absorbed power-law 
model for XBSJ033845.7-352253.

\subsection{XBSJ041108.1--71134}
\begin{figure}
   \centering
   \includegraphics[width=6cm,angle=-90]{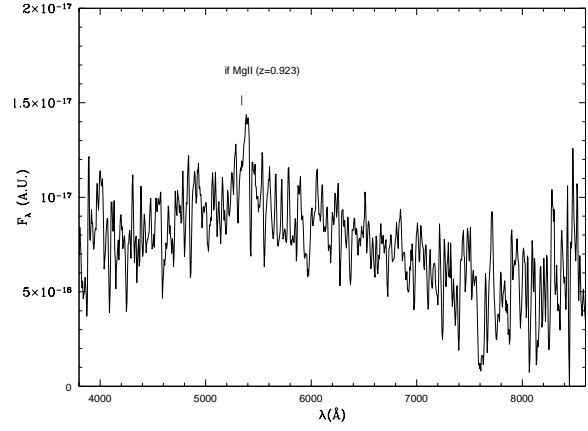}
      \caption{The optical spectrum taken with EFOSC2 in November 2002
of the XBSJ041108.1--71134. The weak (EW=10-15 \AA) emission line
is tentatively identified with MgII$\lambda$2798\AA.
              }
         \label{0411}
   \end{figure}

This is one of the optically faintest sources in the sample (R=20.25).
The noisy spectrum collected in 2400 seconds with EFOSC2  shows only one weak broad 
emission line which we have tentatively identified as the MgII$\lambda$2798\AA\ line. 
With this assumption, the redshift is 0.923.  
The equivalent width of the line is $\sim$10-15\AA\ 
(source rest-frame) which is close to the border-line of the BL Lac objects
classification (EW$<$5\AA, e.g. Stocke et al. 1991). 
The spectrum is reported in Figure~\ref{0411}.
A firmer classification and z estimate requires a better signal-to-noise spectrum.
%

\section{X-ray spectra}
In this appendix we present the spectra of the extragalactic sources (all
AGNs) present in the HBS28 sample.

\begin{figure*}
\centering
\includegraphics[width=18cm]{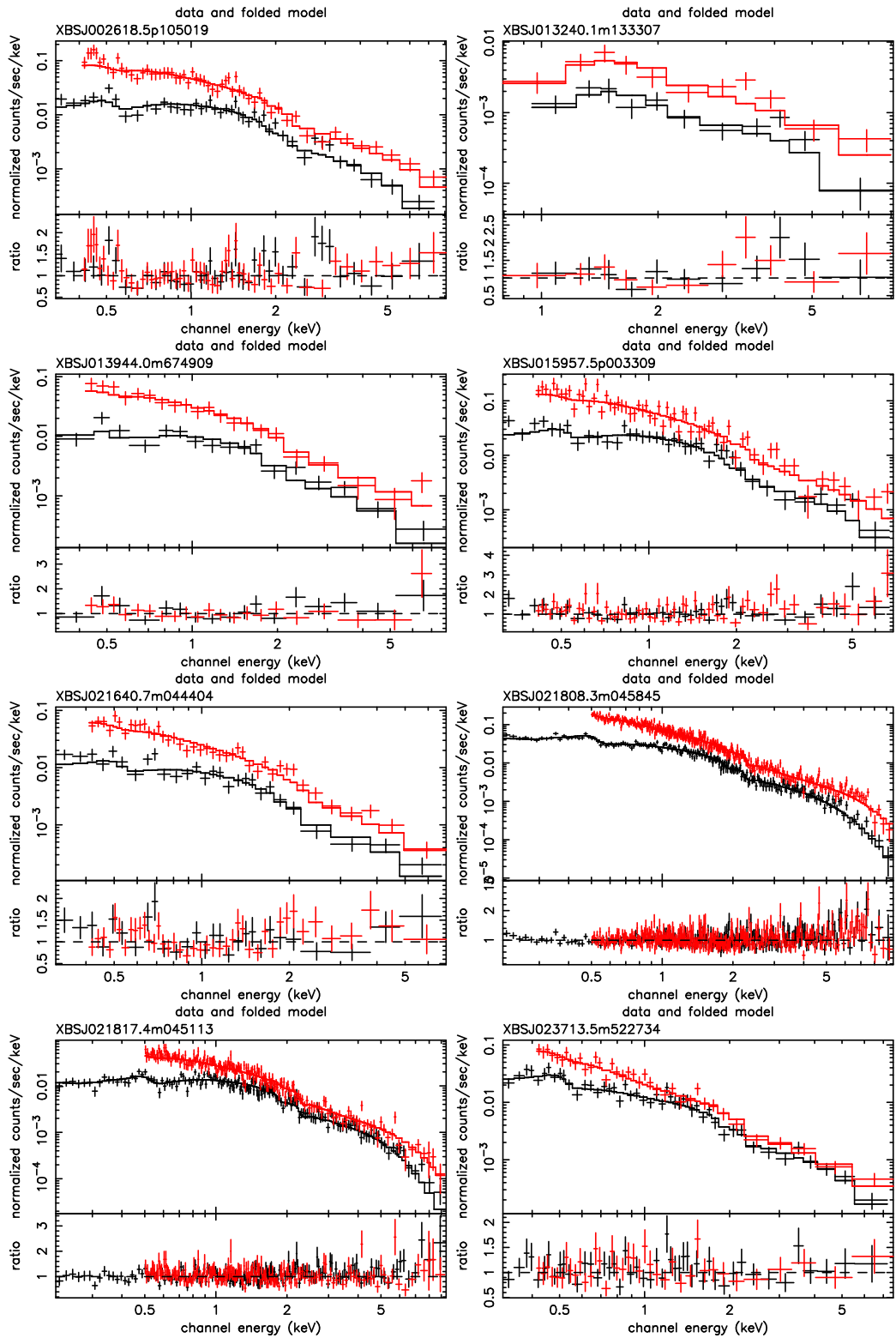}
\caption{The best-fit X-ray spectra of the extragalactic sources in the 
HBS28 sample (note that 
the sources XBSJ021822.2-050615 and XBSJ031859.2--441627 are not included 
since  these spectra have been already presented and discussed in Severgnini 
et al.  2003). 
\label{xspec}}

\end{figure*}
\newpage
\begin{figure*}[h]
\centering
\includegraphics[width=18cm]{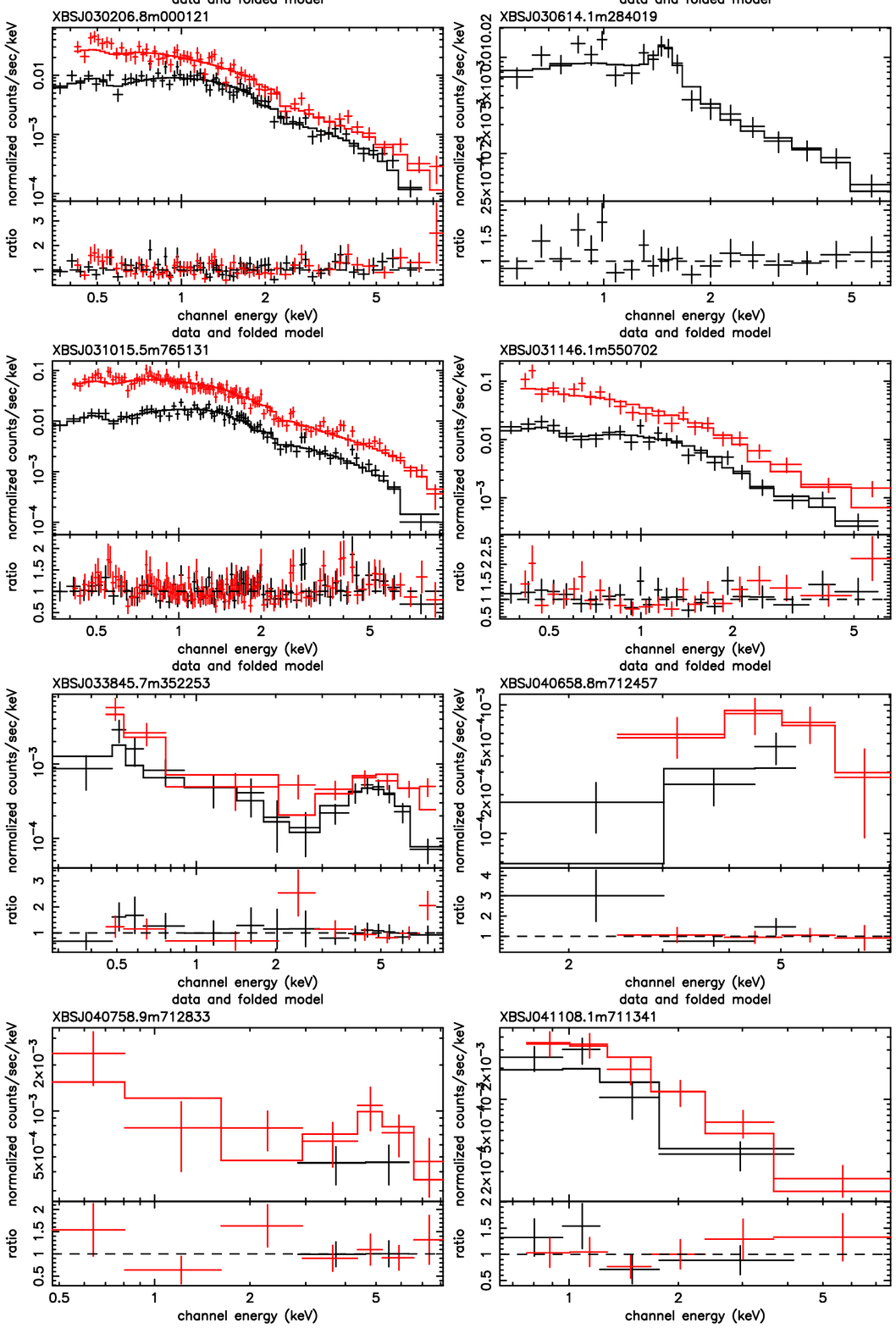}
\addtocounter{figure}{-1}
\caption{(continued)}
\end{figure*}
\newpage
\begin{figure*}[h]
\centering
\includegraphics[width=18cm]{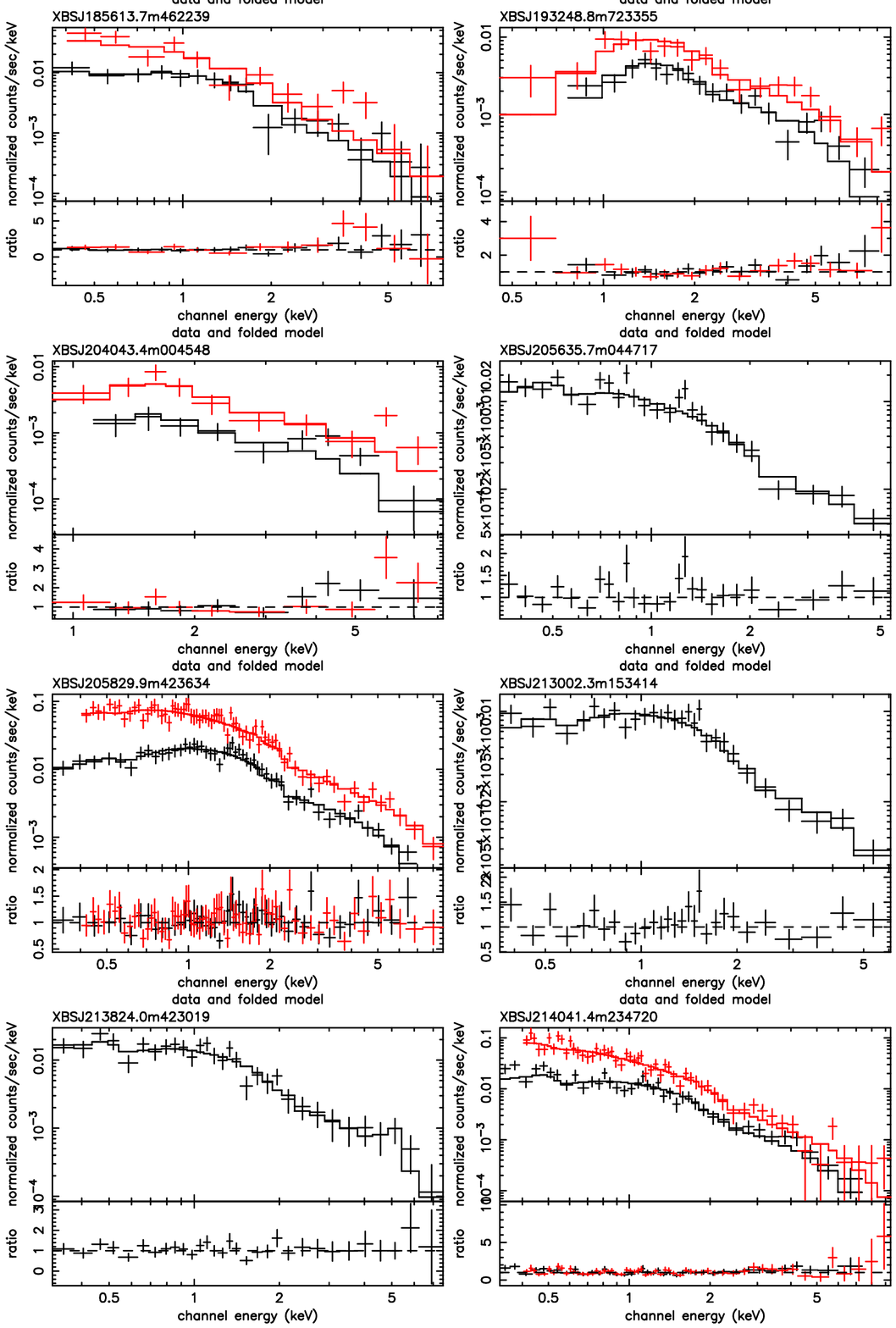}
\addtocounter{figure}{-1}
\caption{(continued)}
\end{figure*}
\newpage
\begin{figure*}[h]
\centering
\includegraphics[width=6cm,angle=-90]{0148fig_b1_d.ps}
\addtocounter{figure}{-1}
\caption{(continued)}
\end{figure*}


\end{document}